\documentclass[ALICE,manyauthors]{cernphprep}

\usepackage[comma,square,numbers,sort&compress]{natbib}
\usepackage{hyperref}
\usepackage[T1]{fontenc}
\usepackage{color}
\usepackage{xspace}
\usepackage{lineno}
\usepackage{enumerate}
\usepackage{orcidlink}

\begin{document}
%

\newcommand{\pp}           {pp\xspace}
\newcommand{\ppbar}        {\mbox{$\mathrm {p\overline{p}}$}\xspace}
\newcommand{\XeXe}         {\mbox{Xe--Xe}\xspace}
\newcommand{\PbPb}         {\mbox{Pb--Pb}\xspace}
\newcommand{\pA}           {\mbox{pA}\xspace}
\newcommand{\pPb}          {\mbox{p--Pb}\xspace}
\newcommand{\AuAu}         {\mbox{Au--Au}\xspace}
\newcommand{\dAu}          {\mbox{d--Au}\xspace}

\newcommand{\s}            {\ensuremath{\sqrt{s}}\xspace}
\newcommand{\snn}          {\ensuremath{\sqrt{s_{\mathrm{NN}}}}\xspace}
\newcommand{\pt}           {\ensuremath{p_{\rm T}}\xspace}
\newcommand{\meanpt}       {$\langle p_{\mathrm{T}}\rangle$\xspace}
\newcommand{\ycms}         {\ensuremath{y_{\rm CMS}}\xspace}
\newcommand{\ylab}         {\ensuremath{y_{\rm lab}}\xspace}
\newcommand{\etarange}[1]  {\mbox{$\left | \eta \right |~<~#1$}}
\newcommand{\yrange}[1]    {\mbox{$\left | y \right |~<~#1$}}
\newcommand{\dndy}         {\ensuremath{\mathrm{d}N_\mathrm{ch}/\mathrm{d}y}\xspace}
\newcommand{\dndeta}       {\ensuremath{\mathrm{d}N_\mathrm{ch}/\mathrm{d}\eta}\xspace}
\newcommand{\avdndeta}     {\ensuremath{\langle\dndeta\rangle}\xspace}
\newcommand{\dNdy}         {\ensuremath{\mathrm{d}N_\mathrm{ch}/\mathrm{d}y}\xspace}
\newcommand{\Npart}        {\ensuremath{N_\mathrm{part}}\xspace}
\newcommand{\Ncoll}        {\ensuremath{N_\mathrm{coll}}\xspace}
\newcommand{\dEdx}         {\ensuremath{\textrm{d}E/\textrm{d}x}\xspace}
\newcommand{\RpPb}         {\ensuremath{R_{\rm pPb}}\xspace}

\newcommand{\nineH}        {$\sqrt{s}~=~0.9$~Te\kern-.1emV\xspace}
\newcommand{\seven}        {$\sqrt{s}~=~7$~Te\kern-.1emV\xspace}
\newcommand{\twoH}         {$\sqrt{s}~=~0.2$~Te\kern-.1emV\xspace}
\newcommand{\twosevensix}  {$\sqrt{s}~=~2.76$~Te\kern-.1emV\xspace}
\newcommand{\five}         {$\sqrt{s}~=~5.02$~Te\kern-.1emV\xspace}
\newcommand{\twosevensixnn}{$\sqrt{s_{\mathrm{NN}}}~=~2.76$~Te\kern-.1emV\xspace}
\newcommand{\fivenn}       {$\sqrt{s_{\mathrm{NN}}}~=~5.02$~Te\kern-.1emV\xspace}
\newcommand{\LT}           {L{\'e}vy-Tsallis\xspace}
\newcommand{\GeVc}         {Ge\kern-.1emV/$c$\xspace}
\newcommand{\MeVc}         {Me\kern-.1emV/$c$\xspace}
\newcommand{\TeV}          {Te\kern-.1emV\xspace}
\newcommand{\GeV}          {Ge\kern-.1emV\xspace}
\newcommand{\MeV}          {Me\kern-.1emV\xspace}
\newcommand{\GeVmass}      {Ge\kern-.2emV/$c^2$\xspace}
\newcommand{\MeVmass}      {Me\kern-.2emV/$c^2$\xspace}
\newcommand{\lumi}         {\ensuremath{\mathcal{L}}\xspace}

\newcommand{\ITS}          {\rm{ITS}\xspace}
\newcommand{\TOF}          {\rm{TOF}\xspace}
\newcommand{\ZDC}          {\rm{ZDC}\xspace}
\newcommand{\ZDCs}         {\rm{ZDCs}\xspace}
\newcommand{\ZNA}          {\rm{ZNA}\xspace}
\newcommand{\ZNC}          {\rm{ZNC}\xspace}
\newcommand{\SPD}          {\rm{SPD}\xspace}
\newcommand{\SDD}          {\rm{SDD}\xspace}
\newcommand{\SSD}          {\rm{SSD}\xspace}
\newcommand{\TPC}          {\rm{TPC}\xspace}
\newcommand{\TRD}          {\rm{TRD}\xspace}
\newcommand{\VZERO}        {\rm{V0}\xspace}
\newcommand{\VZEROA}       {\rm{V0A}\xspace}
\newcommand{\VZEROC}       {\rm{V0C}\xspace}
\newcommand{\Vdecay} 	   {\ensuremath{V^{0}}\xspace}

\newcommand{\ee}           {\ensuremath{e^{+}e^{-}}} 
\newcommand{\pip}          {\ensuremath{\pi^{+}}\xspace}
\newcommand{\pim}          {\ensuremath{\pi^{-}}\xspace}
\newcommand{\kap}          {\ensuremath{\rm{K}^{+}}\xspace}
\newcommand{\kam}          {\ensuremath{\rm{K}^{-}}\xspace}
\newcommand{\pbar}         {\ensuremath{\rm\overline{p}}\xspace}
\newcommand{\kzero}        {\ensuremath{{\rm K}^{0}_{\rm{S}}}\xspace}
\newcommand{\lmb}          {\ensuremath{\Lambda}\xspace}
\newcommand{\almb}         {\ensuremath{\overline{\Lambda}}\xspace}
\newcommand{\Om}           {\ensuremath{\Omega^-}\xspace}
\newcommand{\Mo}           {\ensuremath{\overline{\Omega}^+}\xspace}
\newcommand{\X}            {\ensuremath{\Xi^-}\xspace}
\newcommand{\Ix}           {\ensuremath{\overline{\Xi}^+}\xspace}
\newcommand{\Xis}          {\ensuremath{\Xi^{\pm}}\xspace}
\newcommand{\Oms}          {\ensuremath{\Omega^{\pm}}\xspace}
\newcommand{\degree}       {\ensuremath{^{\rm o}}\xspace}

\definecolor{dgreen}{cmyk}{1.,0.,1.,0.4} 
\definecolor{orange}{cmyk}{0.,0.353,1.,0.} 
\newcommand{\orange}[1]     {\textcolor{orange}{#1}}
\newcommand{\blue}[1]       {{\color{blue}{#1}}}
\newcommand{\green}[1]      {{\color{green}{#1}}}
\newcommand{\dgreen}[1]     {{\textcolor{dgreen}{#1}}}
\newcommand{\red}[1]        {{\color{red}{#1}}}
\newcommand{\magenta}[1]    {{\color{magenta}{#1}}}

\newcommand{\SC}            {\ensuremath{\mathrm{SC}}\xspace}
\newcommand{\NSC}            {\ensuremath{\mathrm{NSC}}\xspace}
\newcommand{\AC}            {\ensuremath{\mathrm{AC}}\xspace}
\newcommand{\NAC}           {\ensuremath{\mathrm{NAC}}\xspace}
\newcommand{\trento}        {T\ensuremath{_\mathrm{R}}ENTo}

\begin{titlepage}
\PHyear{2023}       
\PHnumber{047}      
\PHdate{20 March}  

\title{Higher-order correlations between different moments of two flow amplitudes in \PbPb collisions at $\mathbf{\sqrt{\textit{s}_{\text{NN}}}=5.02}$~\TeV}
\ShortTitle{Higher-order correlations between different moments...}   

\Collaboration{ALICE Collaboration\thanks{See Appendix~\ref{app:collab} for the list of collaboration members}}
\ShortAuthor{ALICE Collaboration} 

\begin{abstract}
The correlations between different moments of two flow amplitudes, extracted with the recently developed asymmetric cumulants, are measured in \PbPb collisions at $\snn = 5.02\;\text{TeV}$ recorded by the ALICE detector at the CERN Large Hadron Collider.
The magnitudes of the measured observables show a dependence on the different moments as well as on the collision centrality, indicating the presence of non-linear response in all even moments up to the eighth.
Furthermore, the higher-order asymmetric cumulants show different signatures than the symmetric and lower-order asymmetric cumulants.
Comparisons with state-of-the-art event generators using two different parametrizations obtained from Bayesian optimization show differences between data and simulations in many of the studied observables, indicating a need for further tuning of the models behind those event generators.
These results provide new and independent constraints on the initial conditions and transport properties of the system created in heavy-ion collisions.

\end{abstract}
\end{titlepage}

\setcounter{page}{2} 

\section{Introduction}
An important ongoing program in high-energy nuclear physics is the exploration of the phase diagram of quantum chromodynamics (QCD) at large values of temperature and/or density. In that regime, quark--gluon plasma (QGP), an extreme state of nuclear matter in which quarks and gluons are deconfined, is formed. Ultrarelativistic heavy-ion collisions at the BNL Relativistic Heavy Ion Collider (RHIC) and the CERN Large Hadron Collider (LHC) recreate the conditions in which the QGP can be produced and, consequently, studied.
Properties of the QGP are typically portrayed by specifying its transport properties (e.g., shear or bulk viscosity), or its equation of state. To date, the QGP remains the state of matter with the smallest ratio of shear viscosity to entropy density ($\eta/s$) ever discovered.
Besides scrutinizing the details of individual states of nuclear matter in the QCD phase diagram, it is equally important to constrain the nature of the transition between different states, i.e., whether it is a smooth crossover, first- or second-order phase transition, etc.
Another extensive effort is allocated to the search for the critical point in the QCD phase diagram, whose existence is still not confirmed experimentally. Recent reviews can be found in Refs.~\cite{Heinz:2013th, Braun-Munzinger:2015hba, Busza:2018rrf, ALICE:2022wpn}.

In non-central heavy-ion collisions, the measured correlations in momentum space among detected particles are dominated by correlations originating from collective anisotropic flow~\cite{Ollitrault:1992bk}.
Since anisotropic flow links the initial-state anisotropies in the collision geometry to the final-state anisotropies in momentum space, its measurements can be used to map out the whole history of the heavy-ion evolution. In particular, anisotropic flow is a sensitive probe of shear viscosity of the QGP, whose collective expansion is successfully described by its hydrodynamic response to the anisotropies in the initial-state geometry.
By using a Fourier series decomposition of the azimuthal distribution $f(\varphi)$ of reconstructed particles, anisotropic flow is quantified with flow magnitudes $v_n$ and symmetry planes $\Psi_n$~\cite{Voloshin:1994mz}
\begin{equation}
f(\varphi)=\frac{1}{2\pi}\left[1+2\sum_{n=1}^{\infty}v_n\cos[n(\varphi-\Psi_n)]\right]\,.
\label{eq:FourierSeries}
\end{equation}
Due to random fluctuations of the impact parameter vector (the vector connecting the centers of the two colliding ions) from one collision to another, this series cannot be obtained directly from the data using the simple relation $v_n e^{in\Psi_n} = \left<e^{in\varphi}\right>$.
Instead, multiparticle azimuthal correlations, which have additional advantages and are not influenced by these fluctuations for suitably chosen harmonics $n_i$, are utilized as experimental estimators according to the following analytic expression~\cite{Bhalerao:2011yg}
\begin{equation}
\left<e^{i(n_1\varphi_1+\cdots+n_k\varphi_k)}\right> = v_{n_1}\cdots v_{n_k}\,e^{i(n_1\Psi_{n_1}+\cdots+n_k\Psi_{n_k})} \,.
\label{eq:generalResult}
\end{equation}
In the above equation, angular brackets indicate an average over all distinct sets of $k$ azimuthal angles of the tracks obtained from the same event. Further technical details about multiparticle azimuthal correlations can be found in Refs.~\cite{Borghini:2000sa, Borghini:2001vi, Bilandzic:2010jr, Bilandzic:2013kga}.

With the multiparticle correlation techniques, the correlations between different flow amplitudes $v_n$ or symmetry planes $\Psi_n$ can be used to extract independent information about all stages in the heavy-ion evolution, when compared to the traditional measurements of individual flow amplitudes $v_n$~\cite{Parkkila:2021yha}.
Among them, the \textit{symmetric cumulants} (SCs) were designed to study the genuine correlations (or cumulants) between the same lowest-order moments of two different flow amplitudes $v_m^2$ and $v_n^2$. 
The simplest realization of SC observables in terms of flow amplitudes and the corresponding experimental estimators built from multiparticle azimuthal correlations is given by~\cite{Bilandzic:2013kga, ALICE:2016kpq, ALICE:2017kwu}
\begin{equation}
\begin{split}
    \SC(m,n) & \equiv \langle v_m^2 v_n^2 \rangle - \langle v_m^2 \rangle\langle v_n^2 \rangle \\
            & = \langle\langle e^{i(m\varphi_1 + n\varphi_2 - m\varphi_3 - n\varphi_4)} \rangle\rangle - \langle\langle e^{im(\varphi_1 - \varphi_2)} \rangle\rangle\langle\langle e^{in(\varphi_1 - \varphi_2)} \rangle\rangle.
\label{eq:SC2h}
\end{split}
\end{equation}
In the first line above, $\langle\cdot\rangle$ indicates an average over all events. Meanwhile, $\langle\langle\cdot\rangle\rangle$ in the second line indicates that the averaging is performed in two steps: first over all distinct particle pairs or quadruplets in an event, and then in the second step the all-event averages are obtained from these single-event averages by weighting them with `number of combinations' multiplicity weights $M(M-1)$ and $M(M-1)(M-2)(M-3)$ for two- and four-particle correlations, respectively~\cite{Bilandzic:2010jr}.
Furthermore, defined this way, SC observables are multivariate cumulants of flow amplitudes squared, and therefore they obey all non-trivial mathematical properties of multivariate cumulants for stochastic variables $v_n^2$ by definition~\cite{doi:10.1143/JPSJ.17.1100}. 
The azimuthal correlators are then used only via Eq.~\eqref{eq:generalResult} to estimate each individual term in the cumulant expansion~\cite{Bilandzic:2021rgb}, i.e., the cumulant expansion is not performed directly on the azimuthal angles.

The particular importance of SCs in flow studies is that they bring new information on QGP properties, as shown in Ref.~\cite{ALICE:2016kpq}.
From experimental measurements of various combinations of $\SC(m,n)$ and their comparison to theoretical predictions, it was concluded that SCs are sensitive to the temperature dependence of $\eta/s$. On the other hand, the analyses of single flow amplitudes can only probe the average value $\langle\eta/s(T)\rangle$~\cite{ALICE:2016kpq, ALICE:2017kwu, ALICE:2021adw}.
Motivated by this access to new information on the system properties, studies were conducted to generalize the SCs from two to any number of harmonics using the cumulant formalism~\cite{Mordasini:2019hut}. The resulting three-harmonic SCs were recently measured by the ALICE Collaboration in \PbPb~collisions recorded in 2010~\cite{ALICE:2021klf} and 2015~\cite{ALICE:2021adw}.

In parallel to the measurements of new flow observables, techniques like Bayesian optimization, where the model parameters are systematically varied to find the best fit to all selected experimental measurements, were applied.
In recent years, multiple Bayesian studies were conducted to constrain the QGP transport properties and details of the initial conditions in \PbPb~collisions~\cite{Bernhard:2015hxa, Bernhard:2016tnd, Bernhard:2019bmu, Auvinen:2020mpc, Nijs:2020ors, Nijs:2020roc, JETSCAPE:2020mzn, Parkkila:2021tqq, Parkkila:2021yha}. While much information on \PbPb~collisions was obtained through Bayesian analyses, the uncertainties on the extracted parameters are still quite large.
As demonstrated in recent studies~\cite{Parkkila:2021tqq, Parkkila:2021yha}, this problem can partially be solved by increasing the scope of the input experimental data, either from various beam-energy ranges or from additional independent observables sensitive to the initial conditions and system properties. 
In recent publications~\cite{Parkkila:2021tqq, Parkkila:2021yha}, dedicated sensitivity studies were also performed on sophisticated higher-harmonic observables for the first time, and it was concluded that the higher-order flow observables exhibit the highest sensitivity to the QGP transport properties.

The success of the SC observables in providing crucial input to Bayesian analyses has prompted a natural quest towards further generalizing the SCs to a new and independent set of multiharmonic observables involving different moments of the flow amplitudes without any influence from the symmetry planes.
Among examples of such estimators are the \textit{asymmetric cumulants} (ACs)~\cite{Bilandzic:2021rgb}. 
As they probe the genuine correlations between different moments of the flow amplitudes, they can, by definition, extract new information (about event-by-event flow fluctuations, non-linear response, etc.) which is inaccessible to the SCs themselves.

The structure of this article is as follows. Section~\ref{sec:theory} reviews the theoretical definitions of ACs and the corresponding experimental estimators. The technical details of the data analysis are introduced in Sec.~\ref{sec:data}. In Sec.~\ref{sec:results}, the first experimental measurements of the ACs and their comparisons with state-of-the-art model predictions are presented. The article is summarized in Sec.~\ref{sec:summary}, while additional support material can be found in the Appendix.

\section{Definitions and scope}
\label{sec:theory}
This article presents the experimental measurements of the AC observables in their simplest realization which involves only two flow amplitudes, evaluated for three different pairs of harmonics chosen to be the same as the ones studied in Ref.~\cite{Parkkila:2021yha}.
For a given pair of different harmonics $(m,n)$, measuring the genuine correlations between $v_m^{2a}$ and $v_n^{2b}$ with AC observables (where $a$ and $b$ are non-equal positive integers) leads to independent information, due to the uniqueness of each multivariate cumulant. Therefore, this analysis focuses on the following new $\AC_{a,b}(m,n)$ flow observables: $\AC_{2,1}(m,n)$, $\AC_{3,1}(m,n)$, and $\AC_{4,1}(m,n)$, to study correlations of the second, third, and fourth moments of $v_m^2$ with the first moment of $v_n^2$.
Two remarks have to be made: the lowest-order AC, $\AC_{1,1}(m,n)$, corresponds to the well-known $\SC(m,n)$ (Eq.~\eqref{eq:SC2h}), showing that ACs are a natural generalization of SCs.
Furthermore, AC observables are invariant under a simultaneous permutation of the indices and the harmonics, e.g., ${\rm AC}_{a,b}(m,n) = {\rm AC}_{b,a}(n,m)$.
These properties will be exploited in Sec.~\ref{sec:results} to discuss the comparisons between $\AC_{a,b}(m,n)$ and $\AC_{a,b}(n,m)$.

The expressions for the ACs in terms of flow amplitudes are derived in Ref.~\cite{Bilandzic:2021rgb} and are recalled as
\begin{equation} \label{eq:AC21-mn}
    \AC_{2,1}(m,n) \equiv \langle v_m^4 v_n^2 \rangle - \langle v_m^4 \rangle\langle v_n^2 \rangle - 2 \langle v_m^2 v_n^2 \rangle\langle v_m^2 \rangle + 2 \langle v_m^2 \rangle^2 \langle v_n^2 \rangle,
\end{equation}
\begin{equation} \label{eq:AC31-mn}
\begin{split}
    \AC_{3,1}(m,n) & \equiv \langle v_m^6 v_n^2 \rangle - \langle v_m^6 \rangle\langle v_n^2 \rangle - 3 \langle v_m^2 v_n^2 \rangle\langle v_m^4 \rangle - 3 \langle v_m^4 v_n^2 \rangle\langle v_m^2 \rangle\\
    & \quad + 6 \langle v_m^4 \rangle\langle v_m^2 \rangle\langle v_n^2 \rangle + 6 \langle v_m^2 v_n^2 \rangle\langle v_m^2 \rangle^2 - 6 \langle v_m^2 \rangle^3 \langle v_n^2 \rangle,
\end{split}
\end{equation}
\begin{equation} \label{eq:AC41-mn}
\begin{split}
    \AC_{4,1}(m,n) & \equiv \langle v_m^8 v_n^2 \rangle - \langle v_m^8 \rangle\langle v_n^2 \rangle - 4 \langle v_m^2 v_n^2 \rangle\langle v_m^6 \rangle - 6 \langle v_m^4 v_n^2 \rangle\langle v_m^4 \rangle + 6 \langle v_m^4 \rangle^2 \langle v_n^2 \rangle - 4 \langle v_m^6 v_n^2 \rangle\langle v_m^2 \rangle\\
    & \quad + 8 \langle v_m^6 \rangle\langle v_m^2 \rangle\langle v_n^2 \rangle + 24 \langle v_m^2 v_n^2 \rangle\langle v_m^4 \rangle\langle v_m^2 \rangle + 12 \langle v_m^4 v_n^2 \rangle\langle v_m^2 \rangle^2 - 36 \langle v_m^4 \rangle\langle v_m^2 \rangle^2 \langle v_n^2 \rangle\\
    & \quad - 24 \langle v_m^2 v_n^2 \rangle\langle v_m^2 \rangle^3 + 24 \langle v_m^2 \rangle^4 \langle v_n^2 \rangle.
\end{split}
\end{equation}
Due to their cumulant nature, much information can already be extracted from the SCs and ACs themselves.
One example is the sign of the measured observables. In Ref.~\cite{ALICE:2016kpq}, it was shown how the event-by-event flow fluctuation patterns and the signatures of SC observables are related. This idea was also applied in the interpretation of the three-harmonic SC presented in Ref.~\cite{ALICE:2021klf}.
For ACs, however, the defining equations are more involved and the direct physical interpretation of their signature, positive or negative, is not straightforward, even in the simplest case of ${\rm AC}_{2,1}$. Another example is the observation of zero magnitude, indicating either an absence of genuine correlations between all involved variables or a trivial consequence of underlying symmetries.
In the case of cumulants, however, this has to be interpreted carefully because in the absence of underlying symmetries either all the cumulants at given order are identically zero, or none of them are identically zero~\cite{doi:10.1143/JPSJ.17.1100}.
Therefore, the nonvanishing results for $\SC(2,4)$, obtained previously in Refs.~\cite{ALICE:2016kpq,ALICE:2017kwu}, imply that all of $\AC_{a,b}(2,4)$ must be nonvanishing as well, and the fact that some of them are consistent with zero would indicate only the limited sensitivity in the available datasets.
In addition, a multivariate cumulant is nonvanishing if and only if all variables in its definition are correlated to each other~\cite{doi:10.1143/JPSJ.17.1100}.

As was done for the SCs, the ACs can also be normalized to remove the sensitivity of the observable to the strength of the individual harmonics and extract only the contribution of the genuine correlations between the various moments of $v_m^2$ and $v_n^2$. This is done according to the standard procedure from Refs.~\cite{Taghavi:2020gcy, Bilandzic:2021rgb}, using
\begin{equation}
    \NAC_{a,1}(m,n) \equiv \frac{\AC_{a,1}(m,n)}{\langle v_m^2 \rangle^a \langle v_n^2 \rangle}, \quad a = 2,\ldots,4.
\label{eq:NAC_mn}
\end{equation}

Equations~\eqref{eq:AC21-mn} to~\eqref{eq:AC41-mn} cannot be used experimentally, as only the azimuthal angles of the final-state particles can be measured. Reliable experimental estimators for the various ACs were presented in Appendix~C.2. of Ref.~\cite{Bilandzic:2021rgb}. As an example, the expression corresponding to Eq.~\eqref{eq:AC21-mn} is
\begin{equation}
\begin{split}
   \AC_{2,1}(m,n) & = \langle\langle {\rm e}^{i(m\varphi_1 + m\varphi_2 + n\varphi_3 - m\varphi_4 - m\varphi_5 - n\varphi_6)} \rangle\rangle\\
    & \quad - \langle\langle {\rm e}^{im(\varphi_1 + \varphi_2 - \varphi_3 -\varphi_4)} \rangle\rangle\langle\langle {\rm e}^{in(\varphi_1 - \varphi_2)} \rangle\rangle \\
    & \quad - 2\langle\langle {\rm e}^{i(m\varphi_1 + n\varphi_2 - m\varphi_3 - n\varphi_4)} \rangle\rangle \langle\langle {\rm e}^{im(\varphi_1 - \varphi_2)} \rangle\rangle \\
    & \quad + 2 \langle\langle {\rm e}^{im(\varphi_1 - \varphi_2)} \rangle\rangle^2 \langle\langle {\rm e}^{in(\varphi_1 - \varphi_2)} \rangle\rangle.
\end{split}
\end{equation}
Each of the azimuthal correlators in the above expression is isotropic (i.e., invariant under the transformation $\varphi\rightarrow\varphi+\alpha$ where $\alpha$ is an arbitrary angle), which ensures that ACs are not affected by random event-by-event fluctuations of the impact parameter vector. In addition, in each azimuthal correlator, each harmonic appears an equal number of times with positive and negative signs, which ensures that any dependence on symmetry planes is canceled out exactly (see Eq.~\eqref{eq:generalResult}).

\section{Data analysis}
\label{sec:data}
The analysis is performed with data from \PbPb collisions at the center-of-mass energy per nucleon pair $\snn = 5.02$~TeV recorded by the ALICE detector in 2015 during the LHC Run 2 period. 

A detailed description of the ALICE detector and its subdetectors can be found in Refs.~\cite{ALICE:2008ngc, ALICE:2014sbx}.
The Time Projection Chamber (TPC)~\cite{Alme:2010ke} was used to reconstruct the charged-particle tracks and measure their momenta.
Two scintillator arrays, V0A and V0C~\cite{ALICE:2008ngc, ALICE:2013axi}, were used for triggering and the determination of the centrality of the collisions~\cite{ALICE:2015juo}. The TPC, V0A, and V0C cover the full azimuthal range within the pseudorapidity ranges $|\eta|<0.9$,\space \space $2.8 < \eta < 5.1$, and $-3.7 < \eta < -1.7$, respectively. 

The detector closest to the beam pipe is the Inner Tracking System (ITS)~\cite{CERN-LHCC-99-012, ALICE:2010tia}, which, besides the TPC, is also used for track reconstruction. It covers the full range in azimuthal angle and consists of six silicon layers, with three different detector technologies.
The two innermost layers, the Silicon Pixel Detectors (SPD), have a high spatial granularity making them ideal for reconstructing both primary and secondary vertices.

Minimum bias (MB) events are triggered by a coincident signal in both the V0A and V0C. The selected MB \PbPb~events all have a reconstructed primary vertex within $\pm 8.0$ cm from the nominal interaction point along the beam direction. 
Background events like beam--gas collisions and pileup were removed using the correlation between the information from the V0 detector and the SPD, as well as the correlation between the number of reconstructed tracks and the information from the ITS and TPC.
A further reduction of the pileup events was obtained using the correlation between the information provided by the SPD and by the Time-Of-Flight detector~\cite{Carnesecchi:2018oss}.
Furthermore, the data taken with each of the two possible directions of the magnetic field in the central barrel are both considered for the analysis.
Finally, an additional event criterion is applied after the track selection to reject all remaining events with less than ten reconstructed tracks. This requirement is motivated by the calculation of the multiparticle azimuthal correlators themselves, and their corresponding event-by-event multiplicity weights. For a $k$-particle correlator, at least $k$ tracks are needed to have a valid expression for the correlator and corresponding multiplicity weight.
This criterion is set to ten tracks, as this is the minimum number of tracks needed to calculate the largest correlator in this analysis, $\AC_{4,1}(m,n)$ (see Appendix~C in Ref.~\cite{Bilandzic:2021rgb} for the full estimator).
After all event selections are applied, approximately 50 million events in the centrality range of $5\%$ to $60\%$ are used in the analysis.
Furthermore, this range is divided into the following centrality classes: the two central classes 5--10\% and 10--20\%, and the four semicentral classes 20--30\%, 30--40\%, 40--50\%, and 50--60\%.

Tracks are reconstructed using the combined information given by the ITS and the TPC.
These tracks are selected to be within $|\eta|<0.8$ and $0.2<\pt<5$ GeV/$c$, where \pt~is the transverse momentum. The low $\pt$ cutoff is implemented for reducing biases due to a smaller reconstruction efficiency at very low transverse momentum, while the high $\pt$ cutoff is chosen for suppressing the contribution to the anisotropies in the measured azimuthal correlations from jets.
In order to avoid contributions from secondary particles, each track is required to have a distance of closest approach (DCA) to the primary vertex of less than 2 cm in the longitudinal direction and less than $0.0105+0.0350/\pt^{1.1}$ cm in the transverse direction, meaning a value ranging from 0.016~cm at \pt $=$ 5.0~GeV/$c$ to 0.22 cm at \pt~$=$~0.2~GeV/$c$.
Additionally, each track is required to have a minimum of 70 TPC space points out of the maximum of 159, and a $\chi^2$ value per space point from the track fit to be within the range of 0.1 to 2.5. The tracks are required to have a minimum of two hits in the ITS.
Requirements were also added to reject tracks with a kink topology (i.e., tracks that appear to change direction due to multiple scatterings and/or weak decays).

Corrections for both non-uniform reconstruction efficiency (NUE) as a function of transverse momentum and non-uniform acceptance (NUA) as a function of azimuthal angle are applied in the same way as in Refs.~\cite{Bilandzic:2013kga, ALICE:2020sup, ALICE:2021klf}. 
For the NUE corrections, a HIJING (Heavy-Ion Jet INteraction Generator)~\cite{Gyulassy:1994ew} simulation with a GEANT3~\cite{Brun:1994aa} detector model is used to correct for the effects of track reconstruction efficiency and contamination from secondary particles by constructing a \pt-dependent track weighting correction. 
The NUA corrections are obtained in a data-driven manner, based on the method presented in Ref.~\cite{Bilandzic:2013kga}.
Finally, the nonflow contribution from the two-particle correlations appearing in the denominator of the NAC is suppressed with the use of a pseudorapidity gap $\vert \Delta\eta \vert > 1$. As the ACs themselves were shown to be robust against nonflow~\cite{Bilandzic:2021rgb}, no gap is used in their computation.

Obtaining the statistical uncertainty through standard error propagation is impractical due to both non-trivial functional dependence and many covariance terms. Therefore, in this analysis, the statistical uncertainties were determined using the standard subsampling procedure (see Appendix~F in Ref.~\cite{Bilandzic:2021rgb}).

The systematic uncertainties are estimated by independent variations of each selection criterion. The significance of each variation is determined according to the Barlow test~\cite{Barlow:2002yb}.
For each variation, the results for $\AC_{a,b}(m,n)$ are compared with the ones obtained with the default selection, taking into account the correlations between their statistical uncertainties.
If the deviation is larger than one $\sigma$, where $\sigma$ is the uncertainty on the difference, the variation is deemed significant according to the Barlow test and is included into the quadratic sum to get the total systematic uncertainties.
Each selection criterion variation is described below and the trend of their relative deviations with respect to the default values observed in semicentral collisions (collisions in the centrality range 20--60\%) is reported in parentheses. Due to the small size of the signal, the variations measured in central events (centrality up to $20$\%) and for higher-order observables can fluctuate more significantly than the ones indicated here.

The effects of the centrality determination are estimated by changing the default V0 estimator to the SPD ($20$\%). The quality of the event selection is further tested by varying the longitudinal position of the primary vertex from $\pm$~8~cm to $\pm$~6~cm ($10$\%), and by applying a tighter pileup rejection factor ($4$\%).
Furthermore, the impact of the two configurations of the magnetic field polarity in the solenoid magnet of ALICE is investigated by performing the same analysis on the data sets taken for each orientation separately ($5$\%).
The minimum number of space points in the TPC required for the track reconstruction is increased from 70 to 80 ($2$\%), and the quality of the reconstruction fit per TPC cluster is reduced from $0.1 < \chi^2 < 2.5$ to $0.1 < \chi^2 < 2.3$ ($4$\%).
The longitudinal DCA is also tightened from 2 cm to 1 cm. It has to be noticed that this systematic variation mainly affects the higher-order combinations (3--60\%).
Finally, a change of the default tracking requirements to the so-called hybrid tracks, which are TPC--ITS tracks with looser DCA requirements in the transverse plane and along the beam axis, leads to an effect on all studied observables (0--30\%).

\section{Results and discussion}
\label{sec:results}
The results obtained in this analysis are first presented together, in order to gain insights on the strength of the different multiharmonic correlations and their dependence on centrality and cumulant order. Then, specific comparisons between the data and the model predictions are presented for a few selected ACs and NACs.

\begin{figure}[t]
    \centering
	\includegraphics[width=0.75\linewidth]{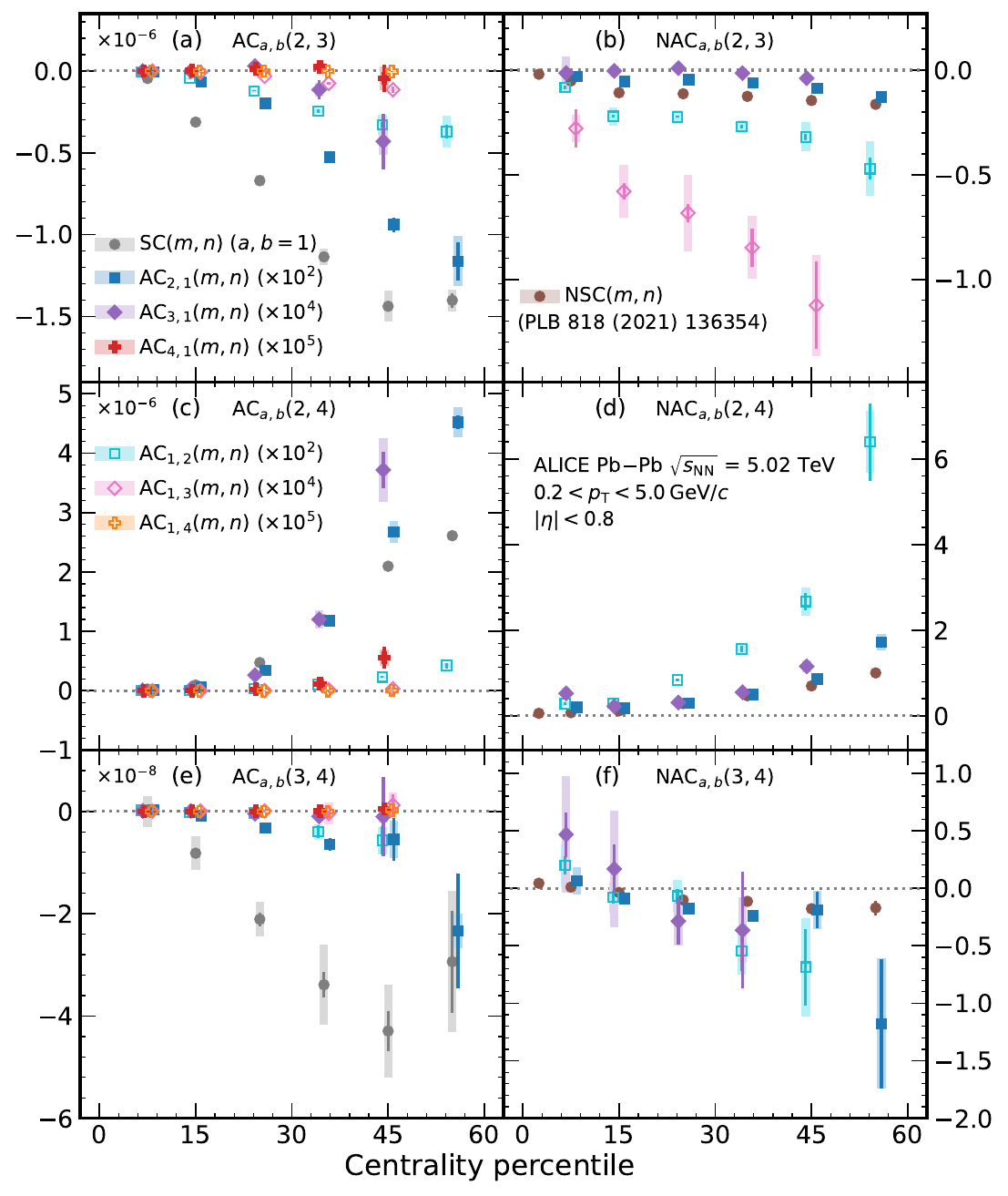}
    \caption{Centrality dependence of the different orders of $\AC_{a,b}(m,n)$ (left) and $\NAC_{a,b}(m,n)$ (right) for four pairs of indices ($a,b$) and three pairs of harmonics ($m,n$) in \PbPb~collisions at \snn~=~5.02~TeV. The mirror combinations of $\AC_{a,b}(m,n)$ (closed markers), i.e., $\AC_{b,a}(m,n)$, are indicated with open markers and similar color scheme. The statistical (systematic) uncertainties are shown with lines (boxes). The data points are shifted horizontally for visibility.}
	\label{fig:ALICEonly}
\end{figure}

The centrality dependence of the thirty-four different $\AC_{a,b}(m,n)$ and $\NAC_{a,b}(m,n)$ is shown in Fig.~\ref{fig:ALICEonly}. However, due to their large uncertainties, $\NAC_{4,1}(m,n)$ and $\NAC_{1,4}(m,n)$ are not presented here.
As the results for $\AC_{a,b}(2,3)$ and $\AC_{a,b}(2,4)$ (Fig.~\ref{fig:ALICEonly}~(a) and~\ref{fig:ALICEonly}~(c), respectively) present similar behaviors, albeit with different signs, they are here discussed together. In the following lines, the notation ($m,n$) thus stands for both pairs of harmonics (2,3) and (2,4).
A dependence of the magnitude of the genuine correlations on the cumulant order is observed. The results for $\SC(m,n) \equiv \AC_{1,1}(m,n)$ present the largest signals, followed by $\AC_{2,1}(m,n)$, while $\AC_{3,1}(m,n)$ and $\AC_{4,1}(m,n)$ are several orders of magnitude smaller.
This decreasing effect is more pronounced for the mirror combinations $\AC_{b,a}(m,n)$ when compared to their respective counterparts.
The splitting between the results for $\AC_{a,b}(m,n)$ and its mirror $\AC_{b,a}(m,n)$ is also present for $\NAC_{a,b}(2,3)$ and $\NAC_{a,b}(2,4)$ (Fig.~\ref{fig:ALICEonly}~(b) and~\ref{fig:ALICEonly}~(d), respectively).
While the results for $\NAC_{2,1}(2,3)$ and $\NAC_{3,1}(2,3)$ demonstrate a trend similar as $\NSC(2,3)$, albeit with smaller magnitudes, the ones for $\NAC_{2,1}(2,4)$ and $\NAC_{3,1}(2,4)$ are in good agreement with $\NSC(2,4)$  and deviate only in more peripheral collisions.
In both panels, the mirror combinations (open markers) show an increase of the magnitude with the cumulant order.
Finally, Figs.~\ref{fig:ALICEonly}~(e) and~\ref{fig:ALICEonly}~(f) present the centrality dependence of the different orders of $\AC_{a,b}(3,4)$ and $\NAC_{a,b}(3,4)$, respectively. The results for $\AC_{a,b}(3,4)$ present the same ordering of the magnitudes with the order of the AC itself, as observed in the two other pairs of harmonics.
However, no splitting between $\AC_{a,1}(3,4)$ and $\AC_{1,a}(3,4)$ can be seen at any order $a = 1,\ldots,4$ within uncertainties.

The measurements are compared with results from simulated events generated with the state-of-the-art chain of models \trento+VISH(2+1)+UrQMD~\cite{Shen:2014vra, Song:2007ux, Bass:1998ca, Bleicher:1999xi}, which will be referred to here as \trento+iEBE-VISHNU.
This model chain consists of the \trento~model~\cite{Moreland:2014oya} for the initial conditions, which are connected by free streaming to a 2+1-dimensional hydrodynamical model VISH(2+1)~\cite{Shen:2014vra}. The description of the heavy-ion evolution after the collision is continued after particlization with a hadronic cascade (UrQMD) model~\cite{Bass:1998ca,Bleicher:1999xi}.
The initial conditions, the specific shear viscosity $\eta/s(T)$, the specific bulk viscosity $\zeta/s(T)$, and other free parameters of this hybrid model, for instance, the Gaussian nucleon width and the minimum inter-nucleon distance, are obtained from a global Bayesian analysis. The data-to-model comparisons are performed with two sets of best-fit parametrizations given by maximum a posteriori (MAP) from two independent Bayesian analysis groups.
As the values for these two sets of best-fit parameters agree within their respective uncertainties from the Bayesian analysis, this work allows the observation of the deviations between the two parametrizations in the predictions of observables that were not used in the Bayesian optimization.
The extraction of the Duke 2019 parametrization~\cite{Bernhard:2019bmu} used a series of ALICE measurements (two-particle correlations, charged-particle multiplicities, etc.) as inputs into the Bayesian analysis. However, most of the used data are from \PbPb~collisions at $\snn = 2.76$~TeV and only a subset of available flow observables were used. 
The Jyv\"{a}skyl\"{a} 2022 parametrization~\cite{Parkkila:2021yha} utilized for the first time higher-order flow observables and data from \PbPb~collisions at both $\snn = 2.76$~TeV and $\snn = 5.02$~TeV.
New observables, such as SC~\cite{ALICE:2016kpq, ALICE:2017kwu, ALICE:2021adw, ALICE:2021klf} and flow harmonic mode couplings~\cite{ALICE:2020sup}, recently measured by the ALICE Collaboration, are more sensitive to the hydrodynamic transport coefficients $\eta/s$ and $\zeta/s$. This resulted in a reduction of the uncertainties of the extracted transport coefficients~\cite{Parkkila:2021yha}.
In these hydrodynamical model calculations, all the kinematic selections, such as \pt~and $\eta$ ranges, are matched with the ones reported in this article. In addition, the observables are extracted from the simulations using the same experimental analysis techniques.

\begin{figure}[b]
	\centering
	\includegraphics[width=0.75\linewidth]{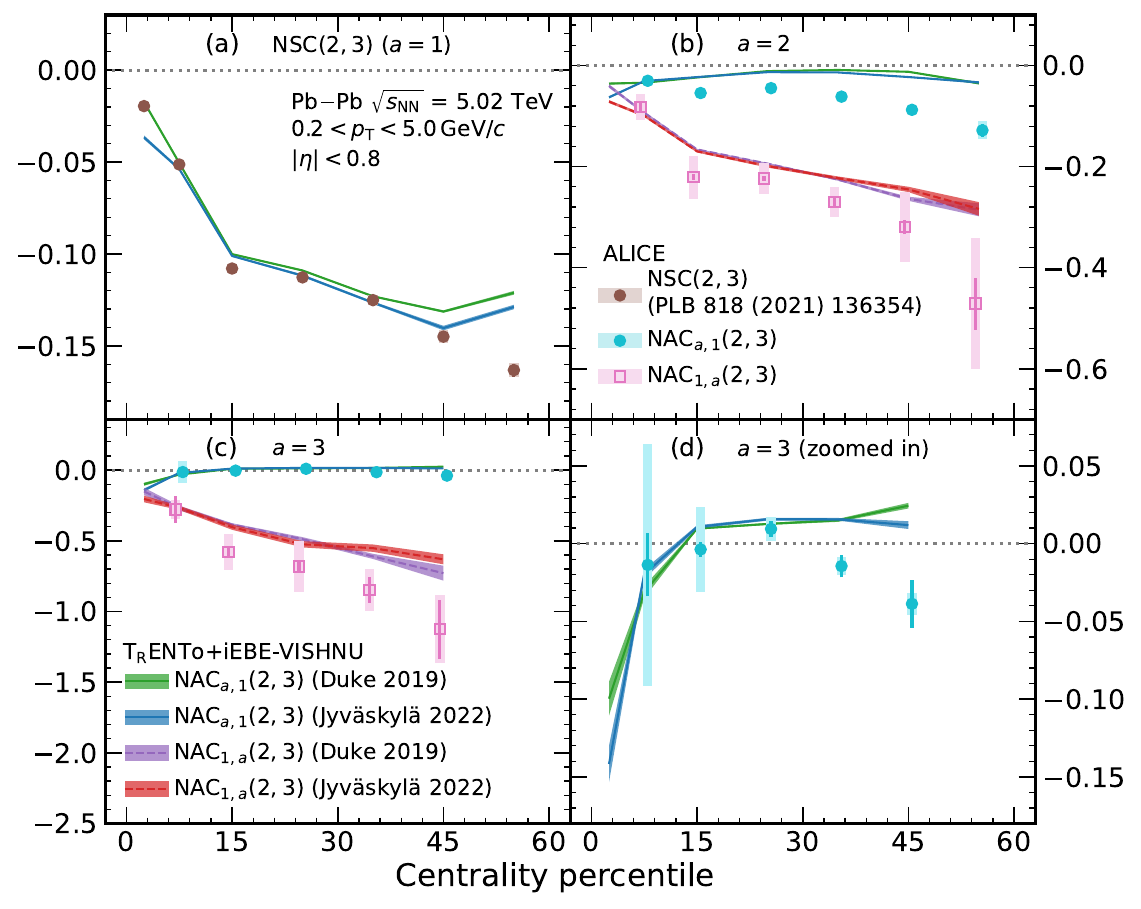}
	\caption{Centrality dependence of the different orders $a = 1,\ldots,3$ of $\NAC_{a,1}(2,3)$ (cyan closed circles) and $\NAC_{1,a}(2,3)$ (pink open squares) compared with the theoretical predictions from \trento+iEBE-VISHNU for the Duke 2019~\cite{Bernhard:2019bmu} and Jyv\"{a}skyl\"{a} 2022~\cite{Parkkila:2021yha} parametrizations. Panel~(d) shows a close-up of the results of $\NAC_{3,1}(2,3)$. The statistical uncertainties in the calculations are indicated by the thicknesses of the colored bands. The data points for $\NSC(2,3)$ are taken from Ref.~\cite{ALICE:2021adw}.}
	\label{fig:NAC23_withModel}
\end{figure}

Figure~\ref{fig:NAC23_withModel} shows the comparisons of the experimental data for $\NAC_{a,b}(2,3)$ with the predictions from hydrodynamic simulation using the Duke 2019 and Jyv\"{a}skyl\"{a} 2022 parametrizations.
As discussed above for the data, the same splitting between $\NAC_{a,1}(2,3)$ and $\NAC_{1,a}(2,3)$ ($a = 2,3)$ can be observed in both model calculations.
The negative sign is visible both for $\NSC(2,3)$ (Fig.~\ref{fig:NAC23_withModel}~(a)) and for some AC observables in the same pair of flow harmonics. The notable exception here is $\NAC_{3,1}(2,3)$, which is close to zero for the data and predicted to be positive in semicentral collisions by both model parametrizations (see Fig.~\ref{fig:NAC23_withModel}~(d) for a close-up of the results).
Furthermore, no major difference can be seen between the predictions from Duke 2019 and Jyv\"{a}skyl\"{a} 2022 for any of the observables in Fig.~\ref{fig:NAC23_withModel}.
As the linear response $v_n = k_n\varepsilon_n\ (n = 2,3)$, where $k_n$ are the linear response coefficients and $\epsilon_n$ the eccentricities in the initial state, dominates in central collisions, the genuine correlations between $v_2$ and $v_3$ are expected to be sensitive only to the initial conditions in this regime. Therefore, a possible explanation of the similar predictions is the use of \trento~for both parametrizations.
Future studies would thus benefit from the comparison between different initial state models.
It also has to be noted that both calculations reproduce the experimental data only qualitatively, and thus the use of such new measurements as input in future Bayesian studies would allow further constraints on the descriptions of the early stages of heavy-ion collisions.

\begin{figure}[b]
	\centering
	\includegraphics[width=0.75\linewidth]{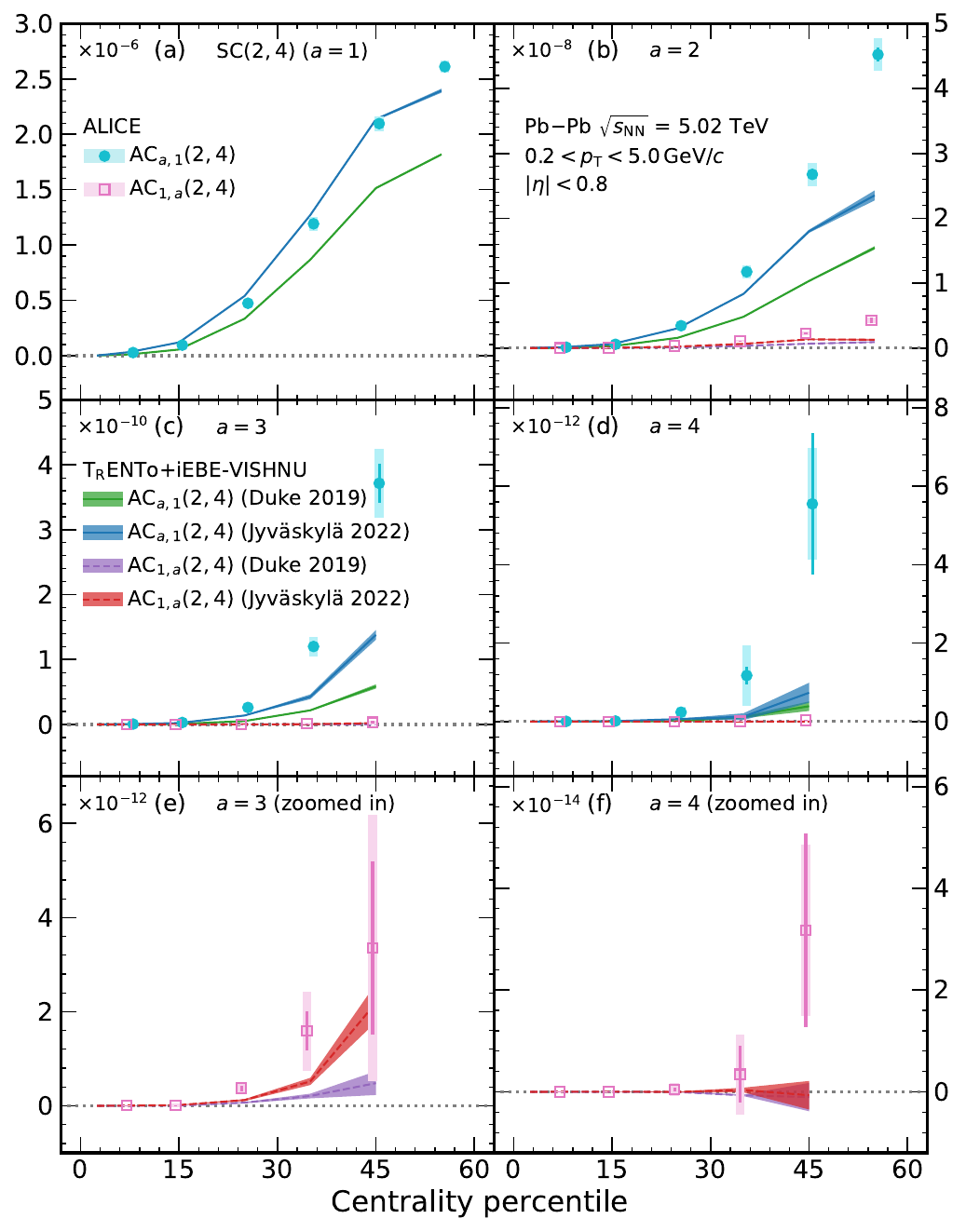}
	\caption{Centrality dependence of the different orders $a = 1,\ldots,4$ of $\AC_{a,1}(2,4)$ (cyan closed circles) and $\AC_{1,a}(2,4)$ (pink open squares) compared with the theoretical predictions from \trento+iEBE-VISHNU for the Duke 2019~\cite{Bernhard:2019bmu} and Jyv\"{a}skyl\"{a} 2022~\cite{Parkkila:2021yha} parametrizations. Panels (e) and (f) show a close-up of the results of $\AC_{1,3}(2,4)$ and $\AC_{1,4}(2,4)$, respectively. The statistical uncertainties in the calculations are indicated by the thicknesses of the colored bands.}
	\label{fig:AC24_withModel}
\end{figure}

The comparison of the results for $\AC_{a,b}(2,4)$ with the calculations from both model parametrizations is shown in Fig.~\ref{fig:AC24_withModel}.
Overall, the behavior of the experimental data is better reproduced by the Jyv\"{a}skyl\"{a} 2022 parametrization, quantitatively for SC and qualitatively for the higher orders. The model predictions are also closer to the experimental values for $\AC_{1,a}(2,4)$ than for $\AC_{a,1}(2,4)$, for $a = 2,3,4$.
The observable $\AC_{a,1}(2,4)$ presents the same positive sign as $\SC(2,4)$. This is the case as well for $\AC_{1,2}(2,4)$ and $\AC_{1,3}(2,4)$ in semicentral collisions (see Fig.~\ref{fig:AC24_withModel}~(e) for a close-up of the latter observable).
However, both experimental and theoretical results for $\AC_{1,4}(2,4)$ are in agreement with zero within uncertainties up to centralities of 30\% for the data and 45\% for the model (Fig.~\ref{fig:AC24_withModel}~(f)).
Because of previous nonvanishing results for $\SC(2,4)$, this indicates that the used sample of Run 2 data is not sufficient to extract nonvanishing values for these particular AC observables.
Past studies have shown the existence of a non-linear response between the fourth-order flow amplitude, $v_4$, and the initial-state ellipticity term $\epsilon_2^2$, which is dominant in non-central collisions~\cite{Gardim:2011xv, Teaney:2012ke}. Furthermore, higher-order harmonics are more sensitive to small variations of $\eta/s$ and $\zeta/s$ than the lower-order coefficients~\cite{Teaney:2012ke, Alver:2010dn, Niemi:2012aj}.
This latter point was further discussed in recent Bayesian estimations~\cite{Parkkila:2021yha}, where higher-order cumulants of flow amplitudes exhibit better sensitivity to QGP properties when compared to the traditional lower-order flow observables, generalizing the findings from Refs.~\cite{Niemi:2012aj, ALICE:2016kpq, ALICE:2017kwu}.
The remaining tensions between data and predictions observed in the current analysis for $\AC_{1,a}(2,4)$ are thus a sign that the measurements of the genuine correlations between $v_2^{2}$ and $v_4^{2a}$ can bring additional input on the higher-order terms of the non-linear response of $v_4$. Comparisons between the initial and final state predictions for these observables would be one of the crucial steps in that direction, as any discrepancies between those would shed more light on the hydrodynamic description of $v_4$ in the models.

\begin{figure}[b]
	\centering
	\includegraphics[width=0.75\linewidth]{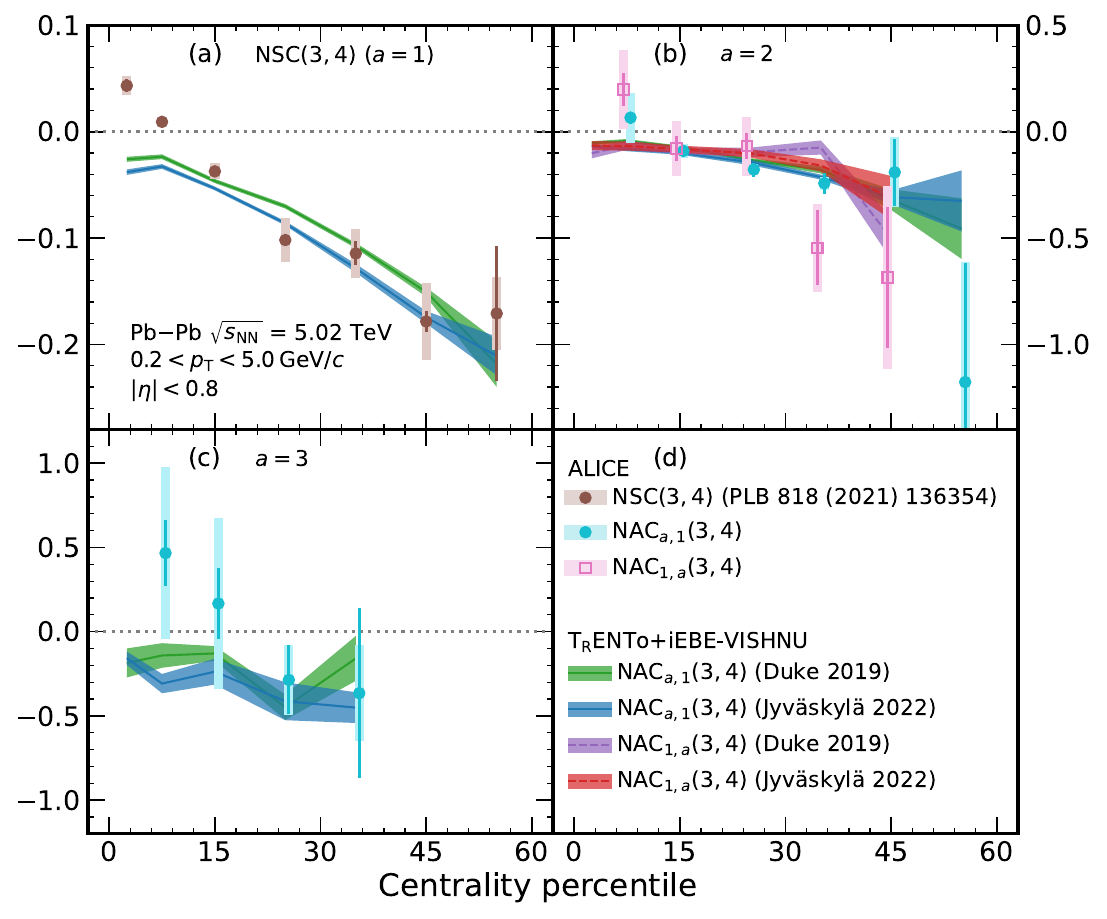}
	\caption{Centrality dependence of the different orders $a = 1,\ldots,3$ of $\NAC_{a,1}(3,4)$ (cyan closed circles) and $\NAC_{1,a}(3,4)$ (pink open squares) compared with the theoretical predictions from \trento+iEBE-VISHNU for the Duke 2019~\cite{Bernhard:2019bmu} and Jyv\"{a}skyl\"{a} 2022~\cite{Parkkila:2021yha} parametrizations. The statistical uncertainties in the calculations are indicated by the thicknesses of the colored bands. The data points for $\NSC(3,4)$ are taken from Ref.~\cite{ALICE:2021adw}.}
	\label{fig:NAC34_withModel}
\end{figure}

The experimental results for $\NAC_{a,b}(3,4)$ are compared with the theoretical calculations from the two model parametrizations, as shown in Fig.~\ref{fig:NAC34_withModel}.
Similarly to the results for the data, the predictions for $\NAC_{2,1}(3,4)$ and $\NAC_{1,2}(3,4)$ are in good agreement with each other within uncertainties (Fig.~\ref{fig:NAC34_withModel}~(b)), for both the Duke 2019 and Jyv\"{a}skyl\"{a} 2022 parametrizations.
Additionally, previous studies~\cite{Gardim:2011xv} have shown that the measurements of $v_4$ can be explained by a contribution from $\epsilon_2^2$ and none from $\epsilon_3$. The observed agreement may then originate from the interplay between the different stages of the evolution of the heavy-ion collisions.
For instance, an agreement between the initial-state predictions for $\NAC_{2,1}(3,4)$ and $\NAC_{1,2}(3,4)$ could indicate an absence of non-linear response between the higher orders of $\epsilon_3$ and $v_4$.
Similarly, different initial-state predictions could mean that the hydrodynamic evolution differently impacts the correlations between the higher moments of $\epsilon_3$ with $v_4$ and the ones between $\epsilon_3$ and the higher moments of $v_4$. However, these non-linear hydrodynamic effects would be such that the two observables converge onto each other in the final state.
Future studies of both the initial and final state, for instance, comparisons of the AC results from this study with initial-state predictions, would help in understanding the origin of the different behaviors --- splitting or no splitting --- observed in the various combinations of NAC.
Furthermore, it is noted that there is good agreement between the predictions obtained from the Duke 2019 and Jyv\"{a}skyl\"{a} 2022 parametrizations, and that they both capture quantitatively the experimental data for centralities higher than 15\%. However, none of the parametrizations capture the sign change between central and semicentral events observed in the experimental data.
In Ref.~\cite{ALICE:2021adw}, this effect was already observed for $\NSC(3,4)$, where the sign change could be reproduced by models using AMPT initial conditions but not by the ones using \trento.
It is therefore expected that future precision measurements of $\NAC_{a,b}(3,4)$ and their use as input data in Bayesian studies would help in tuning the different details of system evolution as a function of the collision centrality in the models.

\begin{figure}[b]
	\centering
	\includegraphics[width=0.75\linewidth]{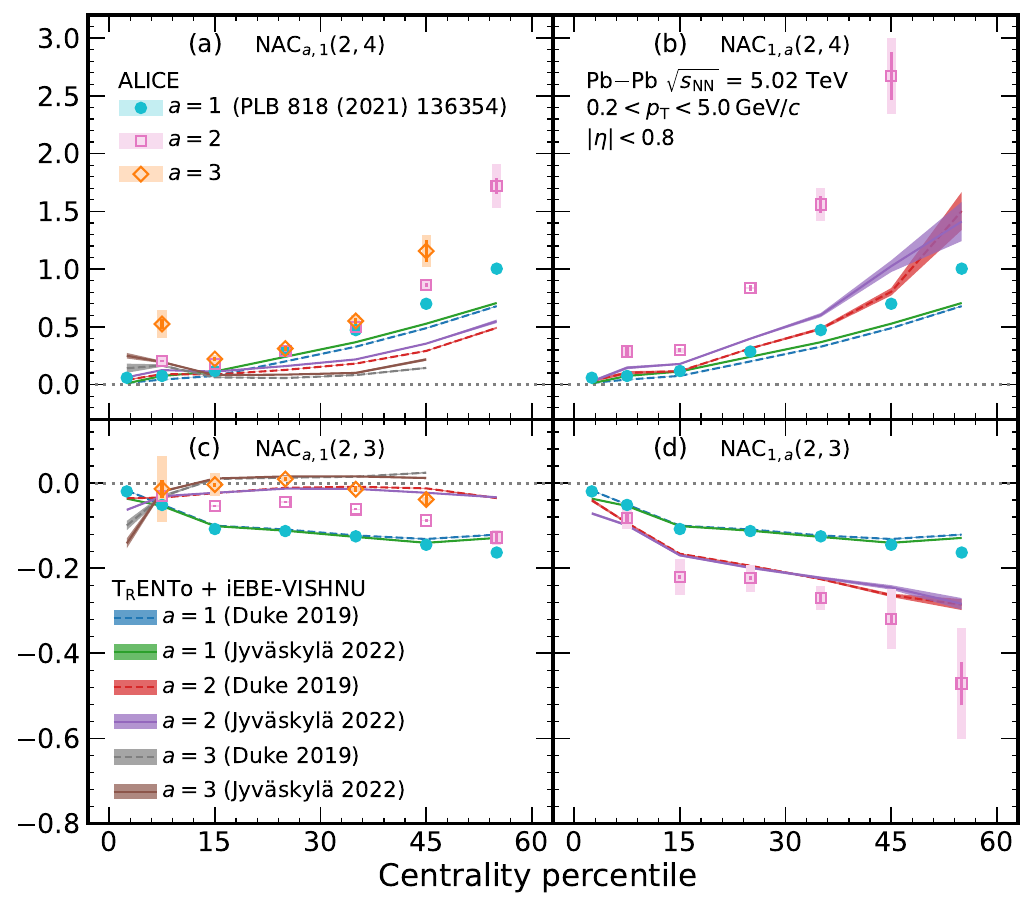}
	\caption{Centrality dependence of the different orders $a = 1,\ldots,3$ of $\NAC_{a,1}(m,n)$ and $\NAC_{1,a}(m,n)$ compared with the theoretical predictions from \trento+iEBE-VISHNU for the Duke 2019~\cite{Bernhard:2019bmu} and Jyv\"{a}skyl\"{a} 2022~\cite{Parkkila:2021yha} parametrizations. The last point of $\NAC_{1,2}(2,4)$ and the results for $\NAC_{1,3}(2,3)$ are not shown for visibility purposes. The statistical uncertainties in the calculations are indicated by the thicknesses of the colored bands. The data points for $a = 1$ are taken from Ref.~\cite{ALICE:2021adw}.}
	\label{fig:NAC_selected}
\end{figure}

Figure~\ref{fig:NAC_selected} summarizes the trends and ordering obtained for $\NAC_{a,b}(2,3)$ and $\NAC_{a,b}(2,4)$.

The main findings of this study on ACs and NACs are summarized here as follows:
\begin{enumerate}[(i)]
	\item The magnitude of the NAC observables varies with different moments as well as with the collision centrality, indicating that non-linear responses are also reflected in higher moments.
	\item While the $\NAC_{2,1}(2,3)$ observable shows a negative decreasing magnitude toward peripheral collisions, the higher-moment correlations, $\NAC_{3,1}(2,3)$, are close to zero except in the most central collisions. The model calculations show a similar trend, but underestimate the data.
	\item Interestingly, the $\AC_{3,1}(2,4)$ observable is much stronger than $\AC_{1,3}(2,4)$ in peripheral collisions. The $\AC_{1,3}(2,4)$ observable is close to zero, while the $\AC_{1,4}(2,4)$ is compatible with zero within uncertainties. As non-zero measurements for SC$(2,4)$~\cite{ALICE:2016kpq} indicate that $v_2^2$ and $v_4^2$ are positively correlated, this does not mean that the higher-order fluctuations of $v_2^2$ and $v_4^2$ are uncorrelated, but that the used data sample is not enough to extract their nonvanishing values.  
	\item $\NAC_{a,1}(2,n)$ shows an inverted dependence on the moments when compared to $\NAC_{1,a}(2,n)$ for $n = 3,4$. No splitting is observed for $\NAC_{a,1}(3,4)$ and $\NAC_{1,a}(3,4)$.
\end{enumerate}

\begin{figure}[b]
	\centering
	\includegraphics[width=0.75\linewidth]{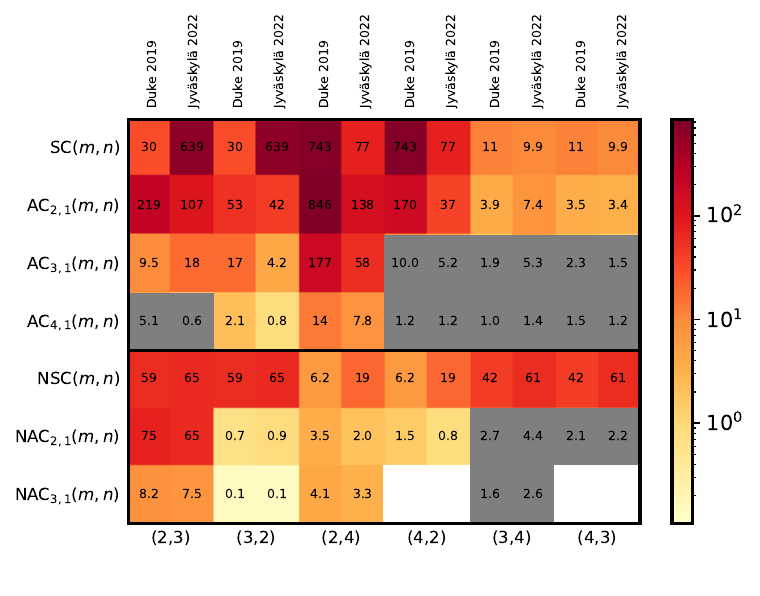}
	\caption{Values for $\chi^2$ test calculated between the data and two different model calculations for all ACs and NACs and all pairs of harmonics presented in this paper. The observables with large uncertainties are indicated with a grey filling, while an absence of observable (see Fig.~\ref{fig:ALICEonly}) is shown with a white filling.}
	\label{fig:chi_squared}
\end{figure}

The agreement between model and data for each cumulant order is now quantified with the $\chi^2$ test ($\chi^2/N_{\rm dof}$), performed as in Ref.~\cite{ALICE:2017kwu}
\begin{equation}
   \chi^2/N_{\rm dof}=\frac{1}{N_{\rm dof}}\sum\limits_{i=1}^{N_{\rm dof}} \frac{(y_{i}-f_{i})^2}{\sigma_{i}^2},
\end{equation}
where $y_{i}$ ($f_{i}$) is a measurement (model) value in a centrality bin ${i}$. The systematic and statistical uncertainties from the data are combined in quadrature $\sigma_{i}^2 = \sigma_{i, \rm stat}^2 + \sigma_{i, \rm syst}^2 + \sigma_{f_{i}, \rm stat}^2$, together with the statistical errors of the model calculations. The quantity $N_{\rm dof}$ corresponds to the number of bins in the centrality range used in $\chi^2$ calculations.
The results are shown in Fig.~\ref{fig:chi_squared} for both model parametrizations indicated on the upper $x$-axis. A grey box indicates that the experimental data have large uncertainties, which may influence the interpretation of the obtained $\chi^2$ value. 
For instance, an agreement between these large uncertainties of the data and the predictions could bias the $\chi^2$ test, leading to an artificially small value. Such bias would then be resolved with a larger data sample.
As they were not measured in the current analysis, $\NAC_{1,3}(2,4)$ and $\NAC_{1,3}(3,4)$ are shown with a white filling. 
A better agreement between the data and a model parametrization is represented by lower $\chi^2$ values.
%
In Fig.~\ref{fig:chi_squared}, it is observed that the model description is better for higher-order harmonics such as all $\AC_{a,b}(3,4)$, and higher-order moments such as $\AC_{4,1}(m,n)$, as well as most of $\NAC_{2,1}(m,n)$ and $\NAC_{3,1}(m,n)$.
Generally, the Jyv\"{a}skyl\"{a} 2022 parametrization describes the data better than the Duke 2019 parametrization. This result is expected as a larger set of data and higher-order observables are included in Jyv\"{a}skyl\"{a} 2022. However, the Duke 2019 still provides better predictions than the newer parametrization in the case of $\SC(2,3)$, $\AC_{3,1}(2,3)$ and $\NSC(m,n)$ for all three pairs of harmonics. 
It is also known that Jyv\"{a}skyl\"{a} 2022 has a deviation in its prediction of $v_2$ at \snn = 5.02 TeV~\cite{Parkkila:2021yha}, which may explain the differences observed in the descriptions of ACs involving $v_2$ but not of the NACs.
Ultimately, the $\chi^2$ test indicates that the independent AC observables are needed as input for future Bayesian analyses in order to reduce the uncertainty on the initial conditions and/or on the hydrodynamic transport coefficients such as $\eta/s(T)$ and $\zeta/s(T)$.

\section{Summary}
\label{sec:summary}
In conclusion, the first measurements of the correlations between different moments of two flow amplitudes in \PbPb~collisions at \snn = 5.02~TeV, using asymmetric cumulants, are shown. This work is a generalization of the previous measurements obtained with symmetric cumulants, proven to have an increased sensitivity to the initial conditions and medium properties.
Due to the cumulant uniqueness, each AC has access to new and independent information unreachable by measuring the SC only. Therefore, this article provides new measurements and resulting additional constraints for 34 new observables.
A dependence of the magnitude of the correlations on the moments and the centrality of the collision is observed, showing an influence of the non-linear response between the higher-moments as well. Furthermore, it has been seen that an increasing (decreasing) dependence with the moments for a given pair $(m,n)$ in $\NAC_{a,1}(m,n)$ generally leads to the decreasing (increasing) dependence for the mirror combination $\NAC_{1,a}(m,n)$.
The comparison of the experimental results with the predictions from two different parametrizations of the \trento+iEBE-VISHNU model chain obtained with Bayesian analyses shows a need for further improvements of the initial conditions and non-linear response. Due to the high sensitivity of these measurements to the model parameters~\cite{Parkkila:2021tqq, Parkkila:2021yha}, the results can improve the current understanding of the detailed nucleon structure in nuclei~\cite{Mantysaari:2016ykx, Mantysaari:2022ffw}, the nuclear skin effect~\cite{Giacalone:2023cet}, and the pre-equilibrium phase~\cite{Romatschke:2017vte, Kurkela:2018wud, Kurkela:2018vqr, Kurkela:2018xxd, Strickland:2018ayk, Kamata:2020mka}. Incorporating these details into the models used by the Duke and Jyv\"{a}skyl\"{a} groups will lead to a better quantitative understanding of the properties of QCD matter. 


\newenvironment{acknowledgement}{\relax}{\relax}
\begin{acknowledgement}
\section*{Acknowledgements}

The ALICE Collaboration would like to thank all its engineers and technicians for their invaluable contributions to the construction of the experiment and the CERN accelerator teams for the outstanding performance of the LHC complex.
The ALICE Collaboration gratefully acknowledges the resources and support provided by all Grid centres and the Worldwide LHC Computing Grid (WLCG) collaboration.
The ALICE Collaboration acknowledges the following funding agencies for their support in building and running the ALICE detector:
A. I. Alikhanyan National Science Laboratory (Yerevan Physics Institute) Foundation (ANSL), State Committee of Science and World Federation of Scientists (WFS), Armenia;
Austrian Academy of Sciences, Austrian Science Fund (FWF): [M 2467-N36] and Nationalstiftung f\"{u}r Forschung, Technologie und Entwicklung, Austria;
Ministry of Communications and High Technologies, National Nuclear Research Center, Azerbaijan;
Conselho Nacional de Desenvolvimento Cient\'{\i}fico e Tecnol\'{o}gico (CNPq), Financiadora de Estudos e Projetos (Finep), Funda\c{c}\~{a}o de Amparo \`{a} Pesquisa do Estado de S\~{a}o Paulo (FAPESP) and Universidade Federal do Rio Grande do Sul (UFRGS), Brazil;
Bulgarian Ministry of Education and Science, within the National Roadmap for Research Infrastructures 2020-2027 (object CERN), Bulgaria;
Ministry of Education of China (MOEC) , Ministry of Science \& Technology of China (MSTC) and National Natural Science Foundation of China (NSFC), China;
Ministry of Science and Education and Croatian Science Foundation, Croatia;
Centro de Aplicaciones Tecnol\'{o}gicas y Desarrollo Nuclear (CEADEN), Cubaenerg\'{\i}a, Cuba;
Ministry of Education, Youth and Sports of the Czech Republic, Czech Republic;
The Danish Council for Independent Research | Natural Sciences, the VILLUM FONDEN and Danish National Research Foundation (DNRF), Denmark;
Helsinki Institute of Physics (HIP), Finland;
Commissariat \`{a} l'Energie Atomique (CEA) and Institut National de Physique Nucl\'{e}aire et de Physique des Particules (IN2P3) and Centre National de la Recherche Scientifique (CNRS), France;
Bundesministerium f\"{u}r Bildung und Forschung (BMBF) and GSI Helmholtzzentrum f\"{u}r Schwerionenforschung GmbH, Germany;
General Secretariat for Research and Technology, Ministry of Education, Research and Religions, Greece;
National Research, Development and Innovation Office, Hungary;
Department of Atomic Energy Government of India (DAE), Department of Science and Technology, Government of India (DST), University Grants Commission, Government of India (UGC) and Council of Scientific and Industrial Research (CSIR), India;
National Research and Innovation Agency - BRIN, Indonesia;
Istituto Nazionale di Fisica Nucleare (INFN), Italy;
Japanese Ministry of Education, Culture, Sports, Science and Technology (MEXT) and Japan Society for the Promotion of Science (JSPS) KAKENHI, Japan;
Consejo Nacional de Ciencia (CONACYT) y Tecnolog\'{i}a, through Fondo de Cooperaci\'{o}n Internacional en Ciencia y Tecnolog\'{i}a (FONCICYT) and Direcci\'{o}n General de Asuntos del Personal Academico (DGAPA), Mexico;
Nederlandse Organisatie voor Wetenschappelijk Onderzoek (NWO), Netherlands;
The Research Council of Norway, Norway;
Commission on Science and Technology for Sustainable Development in the South (COMSATS), Pakistan;
Pontificia Universidad Cat\'{o}lica del Per\'{u}, Peru;
Ministry of Education and Science, National Science Centre and WUT ID-UB, Poland;
Korea Institute of Science and Technology Information and National Research Foundation of Korea (NRF), Republic of Korea;
Ministry of Education and Scientific Research, Institute of Atomic Physics, Ministry of Research and Innovation and Institute of Atomic Physics and University Politehnica of Bucharest, Romania;
Ministry of Education, Science, Research and Sport of the Slovak Republic, Slovakia;
National Research Foundation of South Africa, South Africa;
Swedish Research Council (VR) and Knut \& Alice Wallenberg Foundation (KAW), Sweden;
European Organization for Nuclear Research, Switzerland;
Suranaree University of Technology (SUT), National Science and Technology Development Agency (NSTDA), Thailand Science Research and Innovation (TSRI) and National Science, Research and Innovation Fund (NSRF), Thailand;
Turkish Energy, Nuclear and Mineral Research Agency (TENMAK), Turkey;
National Academy of  Sciences of Ukraine, Ukraine;
Science and Technology Facilities Council (STFC), United Kingdom;
National Science Foundation of the United States of America (NSF) and United States Department of Energy, Office of Nuclear Physics (DOE NP), United States of America.
In addition, individual groups or members have received support from:
European Research Council, Strong 2020 - Horizon 2020 (grant nos. 950692, 824093), European Union;
Academy of Finland (Center of Excellence in Quark Matter) (grant nos. 346327, 346328), Finland;
Programa de Apoyos para la Superaci\'{o}n del Personal Acad\'{e}mico, UNAM, Mexico.

\end{acknowledgement}

\bibliographystyle{utphys}   
\bibliography{bibliography}

\newpage
\appendix
\section{Additional observables}
\label{app:OtherObservables}
The comparisons of the model predictions to the experimental data for $\NAC_{a,b}(2,3)$, $\AC_{a,b}(2,4)$, and $\NAC_{a,b}(3,4)$ are presented in the main part of this article.
In this Appendix, the comparisons for the other observables of this analysis are discussed. The complete list of combinations can be found in Figs.~\ref{fig:ALICEonly} and~\ref{fig:chi_squared}.

\begin{figure}[b]
	\centering
	\includegraphics[width=0.75\linewidth]{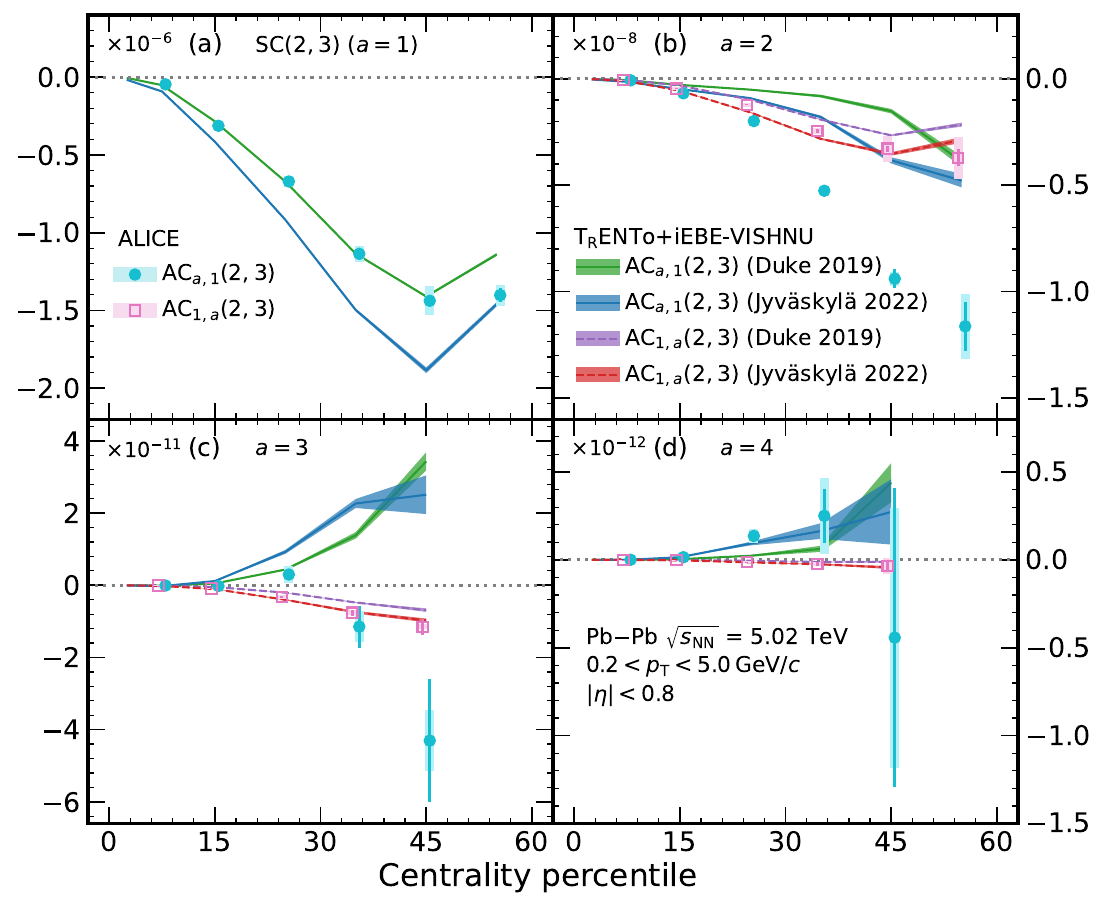}
	\caption{Centrality dependence of the different orders $a = 1,\ldots,4$ of $\AC_{a,1}(2,3)$ (cyan closed circles) and $\AC_{1,a}(2,3)$ (pink open squares) compared with the theoretical predictions from \trento+iEBE-VISHNU for the Duke 2019~\cite{Bernhard:2019bmu} and Jyv\"{a}skyl\"{a} 2022~\cite{Parkkila:2021yha} parametrizations. The statistical uncertainties in the calculations are indicated by the thicknesses of the colored bands.}
	\label{fig:AC23_withModel}
\end{figure}

\begin{figure}[t]
	\centering
	\includegraphics[width=0.75\linewidth]{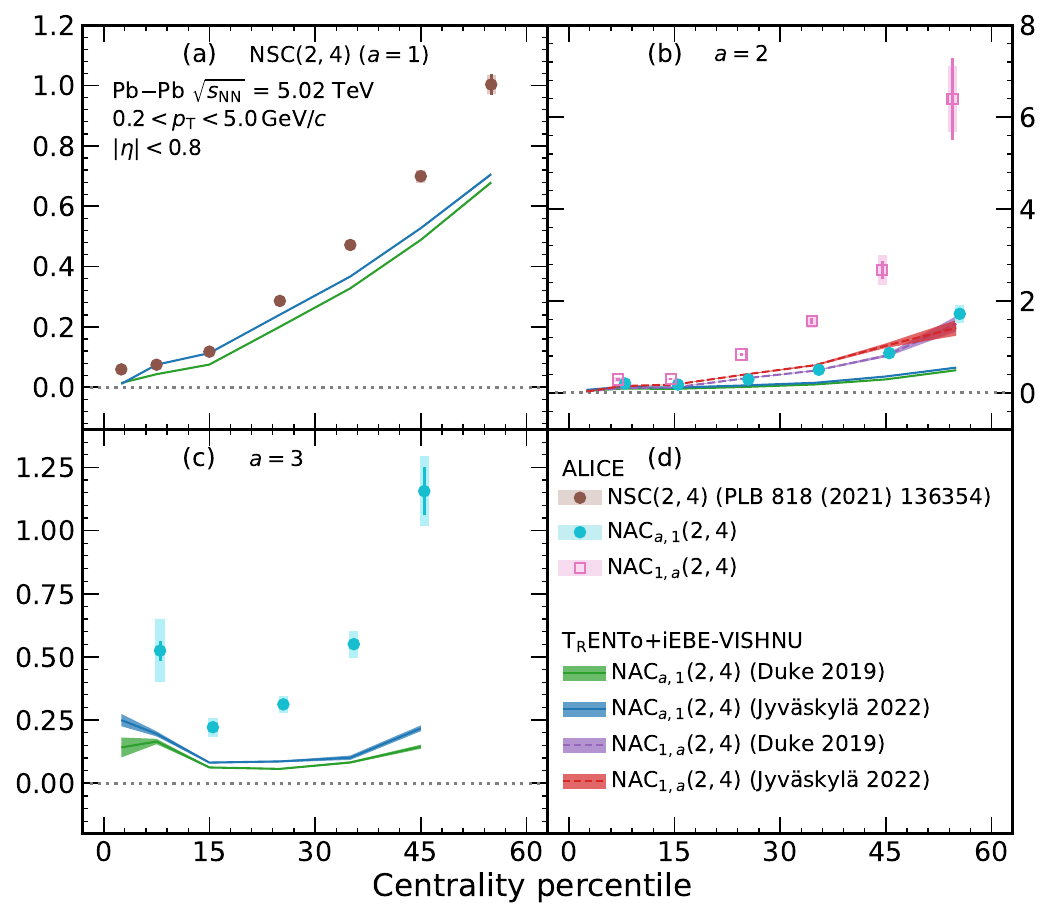}
	\caption{Centrality dependence of the different orders $a = 1,\ldots,3$ of $\NAC_{a,1}(2,4)$ (cyan closed circles) and $\NAC_{1,a}(2,4)$ (pink open squares) compared with the theoretical predictions from \trento+iEBE-VISHNU for the Duke 2019~\cite{Bernhard:2019bmu} and Jyv\"{a}skyl\"{a} 2022~\cite{Parkkila:2021yha} parametrizations. The statistical uncertainties in the calculations are indicated by the thicknesses of the colored bands. The data points for $\NSC(2,4)$ are taken from Ref.~\cite{ALICE:2021adw}.}
	\label{fig:NAC24_withModel}
\end{figure}

The predictions from Duke 2019 and Jyv\"{a}skyl\"{a} 2022 parametrizations and the experimental data for $\AC_{a,b}(2,3)$ and $\NAC_{a,b}(2,4)$ are presented in Figs.~\ref{fig:AC23_withModel} and~\ref{fig:NAC24_withModel}, respectively.
The same splitting with the cumulant order as observed in Fig.~\ref{fig:AC24_withModel} for $\NAC_{a,b}(2,3)$ and $\AC_{a,b}(2,4)$ can be seen here.
In both cases, $\NAC_{a,1}(2,4)$ show increasingly larger magnitudes than $\NAC_{a,1}(4,2)$, while the opposite stands true for the unnormalized AC.
The observables $\AC_{a,b}(2,3)$ and $\AC_{a,b}(2,4)$ (Figs.~\ref{fig:AC23_withModel} and~\ref{fig:AC24_withModel}) show similar agreements between predictions and data. The Jyv\"{a}skyl\"{a} 2022 parametrization has overall a better agreement with the data than the Duke 2019 one, and the higher-order ACs are overall better reproduced than $\SC(m,n)$ and $\AC_{2,1}(m,n)$.
It can be noted, however, that while the Jyv\"{a}skyl\"{a} 2022 parametrization reproduces quantitatively $\SC(2,4)$, whereas the Duke 2019 one does not; the opposite behavior is observed for $\SC(2,3)$.
On the other hand, no big difference in the predictions of $\NAC_{a,1}(2,4)$ and $\NAC_{a,1}(4,2)$ between the Duke 2019 and Jyv\"{a}skyl\"{a} 2022 parametrizations can be observed. Possible effects, such as the ones at the origin of the 10\% discrepancy in the description of $v_2$ at 5.02~TeV by the Jyv\"{a}skyl\"{a} 2022 parametrization~\cite{Parkkila:2021yha}, could be at play in $\SC(2,3)$ and cancel out in $\NSC(2,3)$.

In the case of $\NAC_{a,b}(2,4)$ (Fig.~\ref{fig:NAC24_withModel}), a slightly better agreement with the data from Jyv\"{a}skyl\"{a} 2022 over Duke 2019 parametrizations can be observed. As stated above, the non-linear response between $v_4$ and $\epsilon _2^2$ makes $\AC_{a,b}(2,4)$, and thus $\NAC_{a,b}(2,4)$, especially sensitive to the hydrodynamical description of the collision.

\begin{figure}[t]
	\centering
	\includegraphics[width=0.75\linewidth]{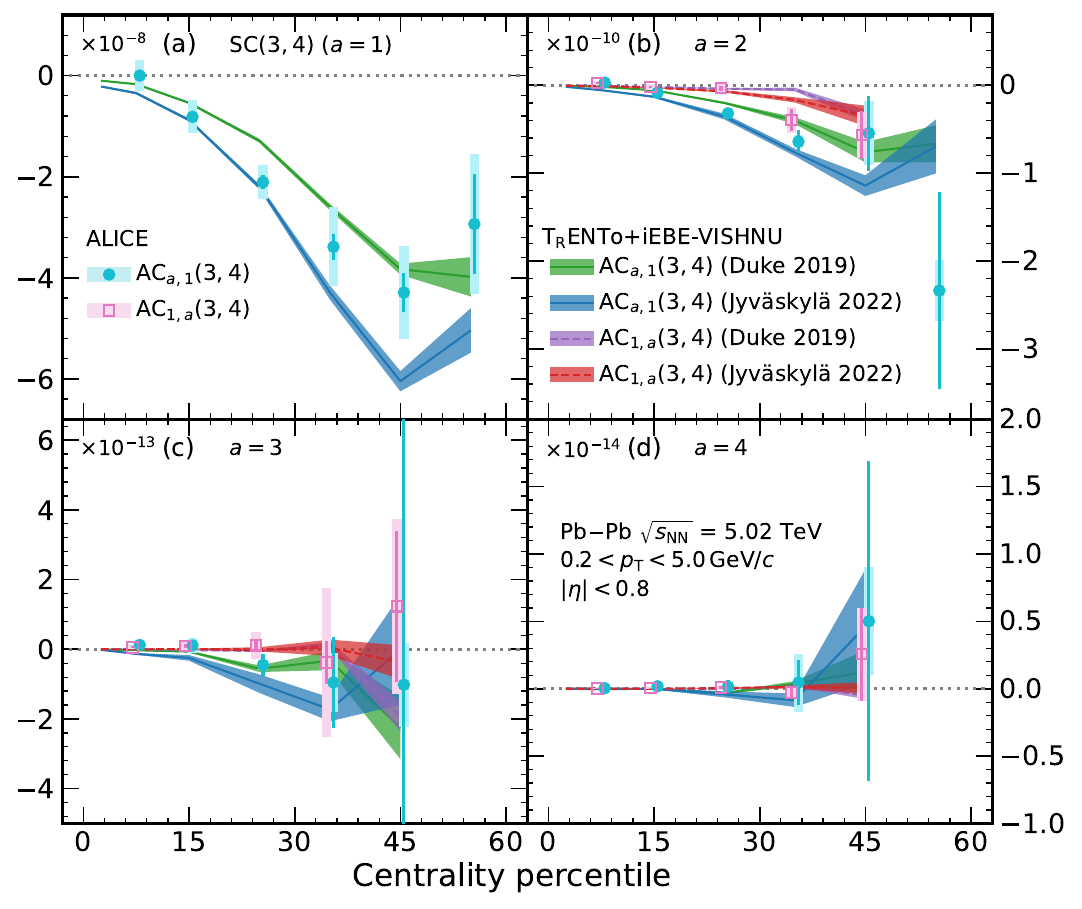}
	\caption{Centrality dependence of the different orders $a = 1,\ldots,4$ of $\AC_{a,1}(3,4)$ (cyan closed circles) and $\AC_{1,a}(3,4)$ (pink open squares) compared with the theoretical predictions from \trento+iEBE-VISHNU for the Duke 2019~\cite{Bernhard:2019bmu} and Jyv\"{a}skyl\"{a} 2022~\cite{Parkkila:2021yha} parametrizations. The statistical uncertainties in the calculations are indicated by the thicknesses of the colored bands.}
	\label{fig:AC34_withModel}
\end{figure}

The predictions for $\AC_{a,b}(3,4)$ are shown in Fig.~\ref{fig:AC34_withModel}.
As was already discussed for $\NAC_{a,b}(3,4)$ (Fig.~\ref{fig:NAC34_withModel}), the measurements for $\AC_{a,1}(3,4)$ and $\AC_{a,1}(4,3)$ are in agreement within uncertainties for both the data and the calculations.
Furthermore, the predictions from the Jyv\"{a}skyl\"{a} 2022 and Duke 2019 parametrizations are both in overall good agreement with the experimental data. 
\clearpage
\newpage

\section{The ALICE Collaboration}
\label{app:collab}
\begin{flushleft} 
\small

S.~Acharya\,\orcidlink{0000-0002-9213-5329}\,$^{\rm 125}$, 
D.~Adamov\'{a}\,\orcidlink{0000-0002-0504-7428}\,$^{\rm 86}$, 
A.~Adler$^{\rm 69}$, 
G.~Aglieri Rinella\,\orcidlink{0000-0002-9611-3696}\,$^{\rm 32}$, 
M.~Agnello\,\orcidlink{0000-0002-0760-5075}\,$^{\rm 29}$, 
N.~Agrawal\,\orcidlink{0000-0003-0348-9836}\,$^{\rm 50}$, 
Z.~Ahammed\,\orcidlink{0000-0001-5241-7412}\,$^{\rm 132}$, 
S.~Ahmad\,\orcidlink{0000-0003-0497-5705}\,$^{\rm 15}$, 
S.U.~Ahn\,\orcidlink{0000-0001-8847-489X}\,$^{\rm 70}$, 
I.~Ahuja\,\orcidlink{0000-0002-4417-1392}\,$^{\rm 37}$
% ...

\section*{Affiliation Notes}

$^{\rm I}$ Deceased\\
$^{\rm II}$ Also at: Max-Planck-Institut f\"{u}r Physik, Munich, Germany\\
$^{\rm III}$ Also at: Italian National Agency for New Technologies, Energy and Sustainable Economic Development (ENEA), Bologna, Italy\\
$^{\rm IV}$ Also at: Department of Applied Physics, Aligarh Muslim University, Aligarh, India\\
$^{\rm V}$ Also at: Institute of Theoretical Physics, University of Wroclaw, Poland\\
$^{\rm VI}$ Also at: An institution covered by a cooperation agreement with CERN\\

\section*{Collaboration Institutes}

$^{1}$ A.I. Alikhanyan National Science Laboratory (Yerevan Physics Institute) Foundation, Yerevan, Armenia\\
$^{2}$ AGH University of Science and Technology, Cracow, Poland\\
$^{3}$ Bogolyubov Institute for Theoretical Physics, National Academy of Sciences of Ukraine, Kiev, Ukraine\\
$^{4}$ Bose Institute, Department of Physics  and Centre for Astroparticle Physics and Space Science (CAPSS), Kolkata, India\\
$^{5}$ California Polytechnic State University, San Luis Obispo, California, United States\\
$^{6}$ Central China Normal University, Wuhan, China\\
$^{7}$ Centro de Aplicaciones Tecnol\'{o}gicas y Desarrollo Nuclear (CEADEN), Havana, Cuba\\
$^{8}$ Centro de Investigaci\'{o}n y de Estudios Avanzados (CINVESTAV), Mexico City and M\'{e}rida, Mexico\\
$^{9}$ Chicago State University, Chicago, Illinois, United States\\
$^{10}$ China Institute of Atomic Energy, Beijing, China\\
$^{11}$ Chungbuk National University, Cheongju, Republic of Korea\\
$^{12}$ Comenius University Bratislava, Faculty of Mathematics, Physics and Informatics, Bratislava, Slovak Republic\\
$^{13}$ COMSATS University Islamabad, Islamabad, Pakistan\\
$^{14}$ Creighton University, Omaha, Nebraska, United States\\
$^{15}$ Department of Physics, Aligarh Muslim University, Aligarh, India\\
$^{16}$ Department of Physics, Pusan National University, Pusan, Republic of Korea\\
$^{17}$ Department of Physics, Sejong University, Seoul, Republic of Korea\\
$^{18}$ Department of Physics, University of California, Berkeley, California, United States\\
$^{19}$ Department of Physics, University of Oslo, Oslo, Norway\\
$^{20}$ Department of Physics and Technology, University of Bergen, Bergen, Norway\\
$^{21}$ Dipartimento di Fisica, Universit\`{a} di Pavia, Pavia, Italy\\
$^{22}$ Dipartimento di Fisica dell'Universit\`{a} and Sezione INFN, Cagliari, Italy\\
$^{23}$ Dipartimento di Fisica dell'Universit\`{a} and Sezione INFN, Trieste, Italy\\
$^{24}$ Dipartimento di Fisica dell'Universit\`{a} and Sezione INFN, Turin, Italy\\
$^{25}$ Dipartimento di Fisica e Astronomia dell'Universit\`{a} and Sezione INFN, Bologna, Italy\\
$^{26}$ Dipartimento di Fisica e Astronomia dell'Universit\`{a} and Sezione INFN, Catania, Italy\\
$^{27}$ Dipartimento di Fisica e Astronomia dell'Universit\`{a} and Sezione INFN, Padova, Italy\\
$^{28}$ Dipartimento di Fisica `E.R.~Caianiello' dell'Universit\`{a} and Gruppo Collegato INFN, Salerno, Italy\\
$^{29}$ Dipartimento DISAT del Politecnico and Sezione INFN, Turin, Italy\\
$^{30}$ Dipartimento di Scienze MIFT, Universit\`{a} di Messina, Messina, Italy\\
$^{31}$ Dipartimento Interateneo di Fisica `M.~Merlin' and Sezione INFN, Bari, Italy\\
$^{32}$ European Organization for Nuclear Research (CERN), Geneva, Switzerland\\
$^{33}$ Faculty of Electrical Engineering, Mechanical Engineering and Naval Architecture, University of Split, Split, Croatia\\
$^{34}$ Faculty of Engineering and Science, Western Norway University of Applied Sciences, Bergen, Norway\\
$^{35}$ Faculty of Nuclear Sciences and Physical Engineering, Czech Technical University in Prague, Prague, Czech Republic\\
$^{36}$ Faculty of Physics, Sofia University, Sofia, Bulgaria\\
$^{37}$ Faculty of Science, P.J.~\v{S}af\'{a}rik University, Ko\v{s}ice, Slovak Republic\\
$^{38}$ Frankfurt Institute for Advanced Studies, Johann Wolfgang Goethe-Universit\"{a}t Frankfurt, Frankfurt, Germany\\
$^{39}$ Fudan University, Shanghai, China\\
$^{40}$ Gangneung-Wonju National University, Gangneung, Republic of Korea\\
$^{41}$ Gauhati University, Department of Physics, Guwahati, India\\
$^{42}$ Helmholtz-Institut f\"{u}r Strahlen- und Kernphysik, Rheinische Friedrich-Wilhelms-Universit\"{a}t Bonn, Bonn, Germany\\
$^{43}$ Helsinki Institute of Physics (HIP), Helsinki, Finland\\
$^{44}$ High Energy Physics Group,  Universidad Aut\'{o}noma de Puebla, Puebla, Mexico\\
$^{45}$ Horia Hulubei National Institute of Physics and Nuclear Engineering, Bucharest, Romania\\
$^{46}$ Indian Institute of Technology Bombay (IIT), Mumbai, India\\
$^{47}$ Indian Institute of Technology Indore, Indore, India\\
$^{48}$ INFN, Laboratori Nazionali di Frascati, Frascati, Italy\\
$^{49}$ INFN, Sezione di Bari, Bari, Italy\\
$^{50}$ INFN, Sezione di Bologna, Bologna, Italy\\
$^{51}$ INFN, Sezione di Cagliari, Cagliari, Italy\\
$^{52}$ INFN, Sezione di Catania, Catania, Italy\\
$^{53}$ INFN, Sezione di Padova, Padova, Italy\\
$^{54}$ INFN, Sezione di Pavia, Pavia, Italy\\
$^{55}$ INFN, Sezione di Torino, Turin, Italy\\
$^{56}$ INFN, Sezione di Trieste, Trieste, Italy\\
$^{57}$ Inha University, Incheon, Republic of Korea\\
$^{58}$ Institute for Gravitational and Subatomic Physics (GRASP), Utrecht University/Nikhef, Utrecht, Netherlands\\
$^{59}$ Institute of Experimental Physics, Slovak Academy of Sciences, Ko\v{s}ice, Slovak Republic\\
$^{60}$ Institute of Physics, Homi Bhabha National Institute, Bhubaneswar, India\\
$^{61}$ Institute of Physics of the Czech Academy of Sciences, Prague, Czech Republic\\
$^{62}$ Institute of Space Science (ISS), Bucharest, Romania\\
$^{63}$ Institut f\"{u}r Kernphysik, Johann Wolfgang Goethe-Universit\"{a}t Frankfurt, Frankfurt, Germany\\
$^{64}$ Instituto de Ciencias Nucleares, Universidad Nacional Aut\'{o}noma de M\'{e}xico, Mexico City, Mexico\\
$^{65}$ Instituto de F\'{i}sica, Universidade Federal do Rio Grande do Sul (UFRGS), Porto Alegre, Brazil\\
$^{66}$ Instituto de F\'{\i}sica, Universidad Nacional Aut\'{o}noma de M\'{e}xico, Mexico City, Mexico\\
$^{67}$ iThemba LABS, National Research Foundation, Somerset West, South Africa\\
$^{68}$ Jeonbuk National University, Jeonju, Republic of Korea\\
$^{69}$ Johann-Wolfgang-Goethe Universit\"{a}t Frankfurt Institut f\"{u}r Informatik, Fachbereich Informatik und Mathematik, Frankfurt, Germany\\
$^{70}$ Korea Institute of Science and Technology Information, Daejeon, Republic of Korea\\
$^{71}$ KTO Karatay University, Konya, Turkey\\
$^{72}$ Laboratoire de Physique des 2 Infinis, Ir\`{e}ne Joliot-Curie, Orsay, France\\
$^{73}$ Laboratoire de Physique Subatomique et de Cosmologie, Universit\'{e} Grenoble-Alpes, CNRS-IN2P3, Grenoble, France\\
$^{74}$ Lawrence Berkeley National Laboratory, Berkeley, California, United States\\
$^{75}$ Lund University Department of Physics, Division of Particle Physics, Lund, Sweden\\
$^{76}$ Nagasaki Institute of Applied Science, Nagasaki, Japan\\
$^{77}$ Nara Women{'}s University (NWU), Nara, Japan\\
$^{78}$ National and Kapodistrian University of Athens, School of Science, Department of Physics , Athens, Greece\\
$^{79}$ National Centre for Nuclear Research, Warsaw, Poland\\
$^{80}$ National Institute of Science Education and Research, Homi Bhabha National Institute, Jatni, India\\
$^{81}$ National Nuclear Research Center, Baku, Azerbaijan\\
$^{82}$ National Research and Innovation Agency - BRIN, Jakarta, Indonesia\\
$^{83}$ Niels Bohr Institute, University of Copenhagen, Copenhagen, Denmark\\
$^{84}$ Nikhef, National institute for subatomic physics, Amsterdam, Netherlands\\
$^{85}$ Nuclear Physics Group, STFC Daresbury Laboratory, Daresbury, United Kingdom\\
$^{86}$ Nuclear Physics Institute of the Czech Academy of Sciences, Husinec-\v{R}e\v{z}, Czech Republic\\
$^{87}$ Oak Ridge National Laboratory, Oak Ridge, Tennessee, United States\\
$^{88}$ Ohio State University, Columbus, Ohio, United States\\
$^{89}$ Physics department, Faculty of science, University of Zagreb, Zagreb, Croatia\\
$^{90}$ Physics Department, Panjab University, Chandigarh, India\\
$^{91}$ Physics Department, University of Jammu, Jammu, India\\
$^{92}$ Physics Program and International Institute for Sustainability with Knotted Chiral Meta Matter (SKCM2), Hiroshima University, Hiroshima, Japan\\
$^{93}$ Physikalisches Institut, Eberhard-Karls-Universit\"{a}t T\"{u}bingen, T\"{u}bingen, Germany\\
$^{94}$ Physikalisches Institut, Ruprecht-Karls-Universit\"{a}t Heidelberg, Heidelberg, Germany\\
$^{95}$ Physik Department, Technische Universit\"{a}t M\"{u}nchen, Munich, Germany\\
$^{96}$ Politecnico di Bari and Sezione INFN, Bari, Italy\\
$^{97}$ Research Division and ExtreMe Matter Institute EMMI, GSI Helmholtzzentrum f\"ur Schwerionenforschung GmbH, Darmstadt, Germany\\
$^{98}$ Saga University, Saga, Japan\\
$^{99}$ Saha Institute of Nuclear Physics, Homi Bhabha National Institute, Kolkata, India\\
$^{100}$ School of Physics and Astronomy, University of Birmingham, Birmingham, United Kingdom\\
$^{101}$ Secci\'{o}n F\'{\i}sica, Departamento de Ciencias, Pontificia Universidad Cat\'{o}lica del Per\'{u}, Lima, Peru\\
$^{102}$ Stefan Meyer Institut f\"{u}r Subatomare Physik (SMI), Vienna, Austria\\
$^{103}$ SUBATECH, IMT Atlantique, Nantes Universit\'{e}, CNRS-IN2P3, Nantes, France\\
$^{104}$ Sungkyunkwan University, Suwon City, Republic of Korea\\
$^{105}$ Suranaree University of Technology, Nakhon Ratchasima, Thailand\\
$^{106}$ Technical University of Ko\v{s}ice, Ko\v{s}ice, Slovak Republic\\
$^{107}$ The Henryk Niewodniczanski Institute of Nuclear Physics, Polish Academy of Sciences, Cracow, Poland\\
$^{108}$ The University of Texas at Austin, Austin, Texas, United States\\
$^{109}$ Universidad Aut\'{o}noma de Sinaloa, Culiac\'{a}n, Mexico\\
$^{110}$ Universidade de S\~{a}o Paulo (USP), S\~{a}o Paulo, Brazil\\
$^{111}$ Universidade Estadual de Campinas (UNICAMP), Campinas, Brazil\\
$^{112}$ Universidade Federal do ABC, Santo Andre, Brazil\\
$^{113}$ University of Cape Town, Cape Town, South Africa\\
$^{114}$ University of Houston, Houston, Texas, United States\\
$^{115}$ University of Jyv\"{a}skyl\"{a}, Jyv\"{a}skyl\"{a}, Finland\\
$^{116}$ University of Kansas, Lawrence, Kansas, United States\\
$^{117}$ University of Liverpool, Liverpool, United Kingdom\\
$^{118}$ University of Science and Technology of China, Hefei, China\\
$^{119}$ University of South-Eastern Norway, Kongsberg, Norway\\
$^{120}$ University of Tennessee, Knoxville, Tennessee, United States\\
$^{121}$ University of the Witwatersrand, Johannesburg, South Africa\\
$^{122}$ University of Tokyo, Tokyo, Japan\\
$^{123}$ University of Tsukuba, Tsukuba, Japan\\
$^{124}$ University Politehnica of Bucharest, Bucharest, Romania\\
$^{125}$ Universit\'{e} Clermont Auvergne, CNRS/IN2P3, LPC, Clermont-Ferrand, France\\
$^{126}$ Universit\'{e} de Lyon, CNRS/IN2P3, Institut de Physique des 2 Infinis de Lyon, Lyon, France\\
$^{127}$ Universit\'{e} de Strasbourg, CNRS, IPHC UMR 7178, F-67000 Strasbourg, France, Strasbourg, France\\
$^{128}$ Universit\'{e} Paris-Saclay Centre d'Etudes de Saclay (CEA), IRFU, D\'{e}partment de Physique Nucl\'{e}aire (DPhN), Saclay, France\\
$^{129}$ Universit\`{a} degli Studi di Foggia, Foggia, Italy\\
$^{130}$ Universit\`{a} del Piemonte Orientale, Vercelli, Italy\\
$^{131}$ Universit\`{a} di Brescia, Brescia, Italy\\
$^{132}$ Variable Energy Cyclotron Centre, Homi Bhabha National Institute, Kolkata, India\\
$^{133}$ Warsaw University of Technology, Warsaw, Poland\\
$^{134}$ Wayne State University, Detroit, Michigan, United States\\
$^{135}$ Westf\"{a}lische Wilhelms-Universit\"{a}t M\"{u}nster, Institut f\"{u}r Kernphysik, M\"{u}nster, Germany\\
$^{136}$ Wigner Research Centre for Physics, Budapest, Hungary\\
$^{137}$ Yale University, New Haven, Connecticut, United States\\
$^{138}$ Yonsei University, Seoul, Republic of Korea\\
$^{139}$  Zentrum  f\"{u}r Technologie und Transfer (ZTT), Worms, Germany\\
$^{140}$ Affiliated with an institute covered by a cooperation agreement with CERN\\
$^{141}$ Affiliated with an international laboratory covered by a cooperation agreement with CERN.\\

\end{flushleft} 
 
\end{document}